\documentclass[a4paper,11pt]{article}
\usepackage{pdfpages}
\usepackage{graphicx}
\usepackage{listings}
\usepackage{caption}
\usepackage{subcaption}
\usepackage{amsmath}
\usepackage{amssymb}
\usepackage{etex,geometry,epsf,amsfonts,cite,microtype,mathtools,pdflscape,hyperref}
\numberwithin{equation}{section}
\hypersetup{colorlinks=true,linkcolor=blue,urlcolor=blue,citecolor=blue}

\newcommand{\sss}{\scriptstyle}
\usepackage{tikz}
\usetikzlibrary{arrows,calc,patterns,decorations.markings,decorations.pathreplacing,plotmarks,shapes.arrows,decorations.pathmorphing,backgrounds}
\tikzset{snake it/.style={decorate, decoration=snake}}

\newcommand\arxiv[2]      {\href{http://arXiv.org/abs/#1}{\tt #2}}
\newcommand\doi[2]        {\href{http://dx.doi.org/#1}{#2}}

\begin{document}


\newcommand{\hq}{{\hat q}}
\newcommand{\htau}{{\hat \tau}}

\newcommand{\pb}[1]{\phi_{#1}}
\newcommand{\psib}{{\bar\psi}}
\newcommand{\hC}{{\hat C}}
\newcommand{\sigmaB}{{\sigma^{}_{\!B}}}
\newcommand{\muB}{{\mu^{}_{\!B}}}

\renewcommand{\vec}[1]{{| #1 \rangle }}
\newcommand{\ket}[1]{{| #1 \rangle }}
\newcommand{\bra}[1]{{\langle {#1}|}}
\newcommand{\kett}[1]{{| #1 \rangle\!\rangle }}
\newcommand{\braa}[1]{{\langle\!\langle {#1}|}}
\newcommand{\kkett}[1]{{\| #1 \rangle\!\rangle }}
\newcommand{\bbraa}[1]{{\langle\!\langle {#1}\|}}
\newcommand{\vac}{{\ket 0}}
\newcommand{\ketbra}[2]{{\ket{#1}\!\!\bra{#2}}}
\newcommand{\kettbraa}[2]{{\kett{#1}\!\braa{#2}}}
\newcommand{\kkettbbraa}[2]{{\kkett{#1}\!\bbraa{#2}}}

\renewcommand{\Re}{{\text{Re}}}
\newcommand{\bpm}{\begin{pmatrix}}
\newcommand{\epm}{\end{pmatrix}}

\newcommand{\NS}{{\text{NS}}}
\newcommand{\R}{{\text{R}}}

\newcommand{\IM}{{\mathrm{Im}}}

\newcommand{\cH}{{\mathcal{H}}}
\newcommand{\cO}{{\mathcal{O}}}

\newcommand{\cHNS}{{\mathcal{H}_\NS}}
\newcommand{\cHR}{{\mathcal{H}_\R}}

\newcommand{\vev}[1]{{\langle\, #1 \,\rangle}}
\newcommand{\Vev}[1]{{\big\langle\, #1 \,\big\rangle}}
\newcommand{\VEV}[1]{{\left\langle\, #1 \,\right\rangle}}
\newcommand{\blank}[1]{}
\newcommand{\bp}{\bar\psi}
\newcommand{\bz}{\bar z}

\newcommand{\rd}{{\mathrm{d}}}
\newcommand{\be}{\begin{equation}}
\newcommand{\ee}{\end{equation}}

\newcommand{\tq}{\tilde q}
\newcommand{\tw}{\tilde w}
\newcommand{\tQ}{\tilde Q}
\newcommand{\tJ}{{\tilde J}}
\newcommand{\ttau}{{\tilde\tau}}

\newcommand{\Zb}{{\mathbb Z}}

\newcommand{\Tr}{{\mathrm{Tr}}}

\thispagestyle{empty}
\rightline{KCL-MTH-21-02}
\begin{center} \vskip 42mm
{\Large\bf 
Free fermions, KdV charges, generalised Gibbs ensembles and modular transforms}\\[10mm]  
{\large 
Max Downing
and
G\'erard M.T.\ Watts
}
\\[5mm]
Department of Mathematics, King's College London,\\
Strand, London WC2R\;2LS, UK
\\[5mm]

\vskip 4mm
\end{center}

\vspace{2cm}
{\centering {\bf Abstract}\\[1mm]}
\begin{quote}

In this paper we consider the modular properties of generalised Gibbs
ensembles in the Ising model, realised as a theory of one free
massless fermion.  
The Gibbs ensembles are given by adding chemical
potentials to chiral charges corresponding to the KdV conserved
quantities. (They can also be thought of as simple models for extended
characters for the W-algebras).  
The eigenvalues and Gibbs ensembles
for the charges can be easily calculated exactly using their
expression as bilinears in the fermion fields.  
We re-derive the
constant term in the charges, previously found by zeta-function
regularisation, from modular properties. 
We expand the Gibbs ensembles as a power series in the chemical potentials and find the
modular properties of the corresponding expectation values of polynomials of KdV
charges.
This leads us to an asymptotic expansion of the Gibbs ensemble
calculated in the opposite channel.
We obtain the same asymptotic
expansion using Dijkgraaf's results for chiral partition functions.
By considering the corresponding TBA calculation, we are led to a
conjecture for the exact closed-form expression of the GGE in the
opposite channel.   
This has the form of a trace over multiple copies of the fermion Fock
space.
We give 
analytic and
%
numerical evidence supporting our conjecture.

\end{quote}

{\setcounter{tocdepth}{2}

}

\newpage
\tableofcontents
\newpage

\section{Introduction}

Two-dimensional conformal field theories have been known for a long time to have an
infinite set of commuting conserved charges.
There is always an infinite set of local charges composed entirely from
modes of the Virasoro algebra that we call ``KdV charges'' for their
equivalence to the charges in the KdV hierarchy \cite{FF1,BLZ1}. There is an
independent infinite set of charges related to the ZMS-Bullough-Dodd
model, see for example \cite{Fring:1992pj}, as well as further (different) sets of charges if the conformal field
theory has an extended chiral  (W-algebra) symmetry.
Most recently, such charges have played a key role in the understanding of
non-thermal or Generalised Gibbs ensembles (GGEs) \cite{GGEs,GGEs2}, and the extension
of the local charges to so-called quasi-local charges \cite{QLC}. These GGEs have already been studied in the large central charge limit in \cite{Dymarsky:2018lhf, Dymarsky:2018iwx}.

One obvious question is that of the modular properties of a GGE. 
The idea of modular invariance of the standard partition function on
the torus, the invariance under reparametrisations of the torus,  was
central to the understanding of conformal field theory and much of the
classification work rests on this. The question is: what are the
modular properties of the partition function of a GGE? 

Independently, there has also been interest in (extended) characters
of W-algebras, both for their relevance for descriptions of Black Holes \cite{KP,GHJ} and in their own right \cite{IW}. 
A W-algebra is a
set of generating
fields amongst which one can find a subset whose zero modes 
$\{W^i_0\}$ commute. It is a natural idea to consider the character
encoding the dimensions of eigenspaces of these zero modes. 
The ``standard'' character includes only
one of the other generators, $L_0$ and its known in certain
circumstances to have well-defined modular properties, and there has
been a natural desire to understand ``extended'' characters
including the full set $\{W^i_0\}$, both from mathematical and
physical considerations. It is to be expected that this problem shares
many of the issues of the GGE and this was in fact one of the main
motivations of this paper, even if the fields in this model do not actually form a W-algebra.

In this paper we look in detail at the simplest
case, that of the KdV charges in the Ising model, or equivalently, the
theory of a single free fermion. In this
model the charges are diagonal on the fermion creation modes and so it is
easy to give an explicit formula for the GGE partition function.
We first give the general setting of the GGE in section \ref{sec:GGE}.
We then describe the free fermion model, its conserved charges and the
construction of the GGE partition 
function in section \ref{sec:2}.
We then investigate its modular properties in 
two
ways: a direct
investigation of the GGE partition function formula in section
\ref{sec:MT1} and
through Dijkgraaf's much earlier work on Chiral deformations of
conformal field theories in 
section \ref{sec:Dijk}.
We arrive at the same results by these two
methods, that the modular transform of a GGE partition function
composed of local charges does not have a convergent expression as a
GGE partition function composed of local charges.
Instead, we find an expression for the modular transform that is
asymptotically correct, 
and 
conjecture a convergent formula for the ground-state contribution.

We also check the modular properties of the expressions we have found
against the modular differential equation results of \cite{Ross}.

We then investigate the same system through the thermodynamic Bethe
ansatz (TBA) in section \ref{sec:TBA}. We find the same expression for
the ground state contribution that we conjectured before, but a richer
set of possible excited state energies. These include the energies
found in the previous sections but with two further sets which give a
correction to the partition function 
with vanishing asymptotic expansion. This leads to a conjectural
formula for the modular transform of the partition function that
includes all sets of ``excited state'' poles from TBA.
In section \ref{sec:numerical} we test this conjecture numerically in
the simplest case and find agreement (within numerical accuracy).

In section \ref{sec:conc},  
we finish with some observations on this formula, its possible proof,
interpretation and extensions.

\section{GGEs}
\label{sec:GGE}

\begin{figure}[htb]
\[
\begin{array}{cccccccccc}
&&
\hfill\mbox{\Large$\llcorner$}\raisebox{1mm}{\kern -1.4mm{$u$}}
&&
\hfill\mbox{\Large$\llcorner$}\raisebox{1mm}{\kern -1.4mm{$v$}}
&&
\hfill\mbox{\Large$\llcorner$}\raisebox{1mm}{\kern -1.4mm{$w$}}
&&
\hfill\mbox{\Large$\llcorner$}\raisebox{1mm}{\kern -1.4mm{$z$}}
\\[-3mm]
\begin{tikzpicture}[baseline=2em]
\begin{scope}[yshift=1.4em]
\draw (0,0.2) -- (2.,0.2);
\draw (0,1.2) -- (2.,1.2);
\draw (0,0.7) ellipse (0.2 and 0.5);
\draw (2.,0.2) arc (-90:90:0.2 and 0.5);
\draw[dashed] (2.,1.2) arc (90:270:0.2 and 0.5);
\draw[->] (-0.48,0.35) arc (-120:150:0.16 and 0.4) node [above left] {$\sss R$};
\draw[<->,dashed] (0,0) -- (2,0) ;
\node at (1,-0.2) {$\sss L$} ;
\fill[black] (2.18,0.5) circle (0.05);
\fill[black] (0.18,0.9) circle (0.05);
\draw[dotted] (0.18,0.5) -- (2.18,0.5);
\draw[->] (0.28,0.5) arc (-30:30:0.16 and 0.4) node [right] {$\sss a$};
\end{scope}
\end{tikzpicture}
&
\raisebox{4mm}{\hbox{$=$}}
&
\begin{tikzpicture}[baseline=2em]
\begin{scope}[yshift=1.2em]
\draw (0,0.2) -- (2.,0.2) -- (2,1.4) -- (0,1.4) -- (0,0.2);
\draw[<->] (0,0) -- (2,0) ;
\draw[<->] (-0.2,0.2) -- (-0.2,1.4);
\node at (1,-0.2) {$\sss L$} ;
\node at (-0.4,0.8) {$\sss R$} ;
\fill[black] (2.,0.2) circle (0.05);
\fill[black] (0.,1.0) circle (0.05);
\draw[->] (0.2,0.2) -- (0.2,0.6) node [right] {$\sss a$}  -- (0.2,1.0);
\end{scope}
\end{tikzpicture}
&
\raisebox{4mm}{\hbox{$\equiv$}}
&
\begin{tikzpicture}[baseline=2em]
\fill[gray!20] (0,0.2) -- (0.8,2.2) -- (0,2.2) -- (0,0.2);
\node at (0.3,1.8) {$\sss A$};
\fill[gray!20] (1.2,0.2) -- (2.0,2.2) -- (1.2,2.2) -- (1.2,0.2);
\node at (1.5,1.8) {$\sss A'$};
\draw (0,0.2) -- (1.2,0.2) -- (1.2,2.2) -- (0,2.2) -- (0,0.2);
\draw[<->] (0,0) -- (1.2,0) ;
\draw[<->] (-0.2,0.2) -- (-0.2,2.2);
\node at (0.6,-0.2) {$\sss R$} ;
\node at (-0.4,1.2) {$\sss L$} ;
\fill[black] (0,0.2) circle (0.05);
\fill[black] (0.8,2.2) circle (0.05);
\draw[->] (0,2.4) -- (0.4,2.4)  node [above] {$\sss a$} -- (0.8,2.4);
\draw[dashed] (0,0.2) -- (0.8,2.2);
\draw (1.2,0.2) -- (2.0,2.2) -- (1.2,2.2);
\draw (0.54,0.3)--(0.66,0.1);
\draw (1.34,2.3)--(1.46,2.1);
\draw (0.33,1.3)--(0.45,1.1);
\draw (0.38,1.3)--(0.5,1.1);
\draw (1.53,1.3)--(1.65,1.1);
\draw (1.58,1.3)--(1.70,1.1);
\end{tikzpicture}
&
\raisebox{4mm}{\hbox{$\equiv$}}
&
\begin{tikzpicture}[baseline=2em]
\draw (0,0.2) -- (1.2,0.2)  node [right] {$\sss 1$};
\fill[black] (0,0.2) circle (0.05);
\fill[black] (0.8,2.2) circle (0.05) node [above] {$\sss\tau$};
\draw (0,0.2) -- (0.8,2.2);
\draw (1.2,0.2) -- (2.0,2.2) -- (0.8,2.2);
\draw (0.54,0.3)--(0.66,0.1);
\draw (1.34,2.3)--(1.46,2.1);
\draw (0.33,1.3)--(0.45,1.1);
\draw (0.38,1.3)--(0.5,1.1);
\draw (1.53,1.3)--(1.65,1.1);
\draw (1.58,1.3)--(1.70,1.1);
\end{tikzpicture}
&
\raisebox{4mm}{\hbox{$\equiv$}}
&
\begin{tikzpicture}[baseline=2em]
\begin{scope}[yshift=3em]
\draw (0,0) circle (0.8);
\draw (0,0) circle (0.3);
\fill[black] (0.8,0) circle (0.05) node [right] {$\sss 1$};;
\fill[black] (0.1,0.28) circle (0.05) node [above] {$\sss q$};;
\draw (-0.38,0.85)--(-0.26,0.65);
\draw (-0.27,0.3)--(-0.15,0.1);
\end{scope}
\end{tikzpicture}
\\
(a) && (b) && (c) && (d) && (e) \end{array}
\]
\caption{(a) The cylinder of length $L$, circumference $R$ with a
  translation through $a$ before the identification of the ends
can be realised as a quotient of the plane (b) and
 (c), and is equivalent (after rescaling) to a
torus (d) with modular parameter $\tau$ and an annulus (e) with coordinates
$w,u,v,z$ respectively. The solid points are identified in each picture.
}
\label{fig1}
\end{figure}
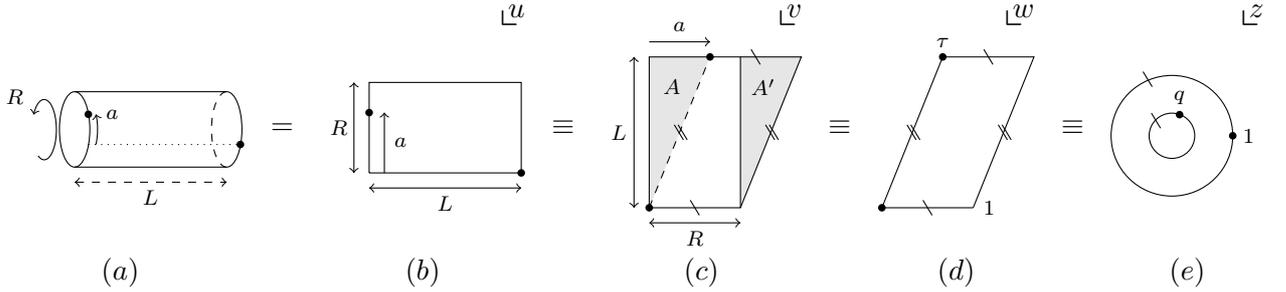

The general setting is of a conformal field theory defined on a
circle
of circumference $R$, with Hamiltonian $H(R)$ and momentum operator $P(R)$. 
We consider the generalised
partition function on a cylinder of length $L$,
\be
  Z(L,R,a) = \Tr( e^{-L H(R)} e^{i a P(R)})
\;,
\label{eq:ZLRa}
\ee
where we have inserted a rotation through a distance $a$ around the
cylinder. This can be calculated as the partition function on the
cylinder where the ends are identified up to a translation at one end,
as in figure \ref{fig1}(a). 

We can realise the cylinder as a quotient of the plane as in figure
\ref{fig1}(b) with a coordinate $u$. We can put this in a different
orientation by mapping to $v = -i u +R$ and, by a rescaling, 
$w = v/R$, this is equivalent to a torus with modular parameter 
$\tau = i(L/R) +(a/R)$ and by a conformal map $z=\exp(2\pi iw)$, to an annulus.

To evaluate the partition function \eqref{eq:ZLRa}, we map the
cylinder to the plane to find
\be
 H(R) = \frac{2\pi}{R}(L_0 + \bar L_0 - \tfrac c{12})
\;,\;\;
 P(R) = \frac{2\pi}{R}(L_0 - \bar L_0)
\;,
\ee
with the result that
\be
  Z(L,R,a) \equiv Z(\tau,\bar\tau)
= \Tr( q^{L_0 - \tfrac c{24}}\,\bar q^{\bar L_0 - \tfrac c{24}} )
\;,\;\;
  q = e^{2\pi i \tau}
\;.
\label{eq:tr2}
\ee
\subsection{Modular invariance}
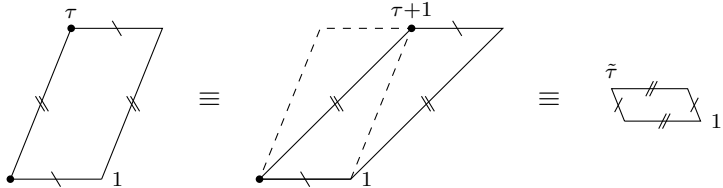
\begin{figure}[htb]
\[
\begin{array}{ccccc}
\begin{tikzpicture}[baseline=2em]
\draw (0,0.2) -- (1.2,0.2)  node [right] {$\sss 1$};
\fill[black] (0,0.2) circle (0.05);
\fill[black] (0.8,2.2) circle (0.05) node [above] {$\sss\tau$};
\draw (0,0.2) -- (0.8,2.2);
\draw (1.2,0.2) -- (2.0,2.2) -- (0.8,2.2);
\draw (0.54,0.3)--(0.66,0.1);
\draw (1.34,2.3)--(1.46,2.1);
\draw (0.33,1.3)--(0.45,1.1);
\draw (0.38,1.3)--(0.5,1.1);
\draw (1.53,1.3)--(1.65,1.1);
\draw (1.58,1.3)--(1.70,1.1);
\end{tikzpicture}
&\;\raisebox{4mm}{\mbox{$\equiv$}}\;
&
\begin{tikzpicture}[baseline=2em]
\draw (0,0.2) -- (1.2,0.2)  node [right] {$\sss 1$};
\fill[black] (0,0.2) circle (0.05);
\fill[black] (2.0,2.2) circle (0.05) node [above] {$\sss\tau+1$};
\draw[dashed] (1.2,0.2) -- (2.0,2.2) -- (0.8,2.2) -- (0,0.2);
\draw (0,0.2)--(1.2,0.2) -- (3.2,2.2) -- (2.0,2.2)--(0,0.2);
\draw (0.54,0.3)--(0.66,0.1);
\draw (2.54,2.3)--(2.66,2.1);
\draw (0.93,1.3)--(1.05,1.1);
\draw (0.98,1.3)--(1.1,1.1);
\draw (2.13,1.3)--(2.25,1.1);
\draw (2.18,1.3)--(2.30,1.1);
\end{tikzpicture}
&\;\raisebox{4mm}{\mbox{$\equiv$}}\;
&
\begin{tikzpicture}[baseline=2em]
\begin{scope}[yshift=2em]
\draw (0,0.2) -- (1,0.2) node [right] {$\sss 1$}  -- (0.83,0.63) -- (-.17,.63) node  [above] {$\sss\tilde\tau$}-- (0,0.2);
\draw (0.45,0.1)--(0.55,0.3);
\draw (0.5,0.1)--(0.6,0.3);
\draw (0.28,0.53)--(0.38,0.73);
\draw (0.33,0.53)--(0.43,0.73);
\draw (-0.13,0.32)--(-0.03,0.51);
\draw (1-0.13,0.32)--(1-0.03,0.51);
\end{scope}
\end{tikzpicture}
\end{array}
\]
\caption{\parbox[t]{.6\textwidth}{The torus with parameter $\tau$ is equivalent to the tori
with parameters $\tau+1$ and  $\ttau=-1/\tau$.}}
\label{fig3}
\end{figure}

The parametrisation of the torus by the complex parameter $\tau$ is
not unique - tori with parameters $\tau$ and
$\tilde\tau=(a\tau+b)/(c\tau+d)$ related by the action of the modular
group $SL(2,\mathbb Z)$ 
are conformally equivalent.
The modular group is generated by 
$T:\tau
\mapsto 1+\tau$ and $S:\tau \mapsto -1/\tau$, as shown in 
figure \ref{fig3}. The equivalence under $S$ corresponds to the map $w \mapsto
\tilde w = 1 - w/\tau$.

Modular invariance is the statement that the partition functions
evaluated for conformally equivalent tori are equal,
\be
 Z(\tau,\bar\tau) = Z\left(\frac{a\tau + b}{c\tau + d},\frac{a\bar\tau + b}{c\bar\tau + d}\right)
\;.
\label{eq:modinv0}
\ee
If a theory contains a conserved charge $Q$, we can consider a
modified trace including the charge $Q$ with a chemical potential, and
ask whether this can also be expressed as a trace after a modular
transformation under which $q \to \tq = \exp(-2\pi i/\tau)$ and $Q \to \tilde Q$,
\be
Z(\tau,\bar\tau,Q)=
\Tr( q^{L_0 - \tfrac c{24}}\,\bar q^{\bar L_0 - \tfrac c{24}} \;e^Q\; )
\overset{?}{=} 
\Tr( \tq^{L_0 - \tfrac c{24}}\,\bar \tq^{\bar L_0 - \tfrac c{24}}
\;e^{\tQ}\; )
=Z(\ttau,\tilde{\bar\tau},\tQ)
\;.
\label{eq:GGEmt}
\ee 
This charge $Q$ could represent a GGE composed of a discrete set of higher spin
charges $Q_i$ or a continuous set of quasi-local charges $Q(t)$,
\be Q = \sum_i \alpha_i Q_i + \int \alpha(t) Q(t) \,\rd t
\;,\label{eq:Q decomposition}
\ee
or the generator of a W-algebra
\be
  Q = \alpha W_0\
\;.
\ee
In this paper we will focus on the former case and whether we can make
any sense of equation \eqref{eq:GGEmt} in the free fermion model.

\subsection{Conserved charges}

The simplest charges are those which are the integral of a local
holomorphic current $J$.
We can always choose $J(w)$ to be a quasi-primary field of weight $h$
(roughly speaking a non-derivative field) which transforms under the
M\"obius group action $w \mapsto w' = (aw+b)/(cw+d)$ 
as
\be
  J(w) = \left(\frac{\partial w'}{\partial w}\right)^h\, J(w')
\;.
\ee
The map from the cylinder to the torus, $z = \exp(2\pi w/R)$ is not,
however, a M\"obius map and we need to consider how the
quasi-primary current $J(w)$ transforms. The rule for the
stress-energy current is well-known, 
\be
  T(w) = \left(\frac{2\pi}{R}\right)^2\left( z^2 T(z) - \tfrac c{24}
  \right)
\;.
\label{eq:Txform}
\ee
In the general case, Gaberdiel has given a method to find the
transformation law of a general quasi-primary field, \cite{Gaberdiel}.

It turns out that it is the charges defined as integrals of
quasi-primary fields on the torus that have simple transformation
properties, for the following reason.
It is straightforward to show that the
expectation value $\vev{ Q }_\tau$ of the charge
\be 
Q = \int_0^1 J(x) \,\rd x
\;,
\ee
on the torus with periods 1 and $\tau$, as in figure \ref{fig3}, 
is not invariant but is instead a modular form of weight $h$.  
If $\ttau = (a\tau + b)/(c\tau + d)$, then a modular form $f(\tau)$ of
weight $h$ transforms as
$f(\ttau) = (c\tau + d)^h f(\tau)$, see appendix \ref{app:mf} for a more detailed discussion of modular forms.
To prove that $\vev{Q}_\tau$ satisfies this law, we only need to
consider the behaviour under the generators $S$ and 
$T$  of the modular group. 

The first step is to relate the integral along the real axis to the
integral over the torus, using translation invariance of the
expectation value on the torus:
\be
\Vev{Q}_\tau
= \Vev{\int_0^1 J(x) \,\rd x}_\tau
= \Vev{ \iint J(w) \,\rd x \,\frac{\rd y}{\IM\,\tau}}_\tau
= \Vev{ \iint J(w) \,\frac{\rd^2 w}{2i\,\IM\,\tau}}_\tau
\;.
\label{eq:ev1}\ee
This is clearly invariant under $T$ since $\langle J(w)\rangle_\tau$ can be extended from the fundamental domain of the torus to the whole plane by $\langle J(w)\rangle_\tau = \langle J(w+m+n\tau)\rangle_\tau$ for $m,n\in\mathbb{Z}$, and so the integrals over region A and A$^\prime$ in figure \ref{fig1} are equal:
\be
  \Vev{Q}_\tau = \Vev{Q}_{\tau+1}
\;.
\label{eq:Txfmn}
\ee
For $S:\tau\mapsto\ttau=-1/\tau$, we consider the M\"obius map $\tilde w
\mapsto w = 1-\tilde w/\ttau = 1 + \tilde w \tau$.
Using the fact that the measure $\rd^2 w/\IM\,\tau$ is invariant
under the M\"obius group, and $J$ is a quasi-primary field of weight $h$, we get
\be
\Vev{Q}_\ttau
=
\Vev{ \iint J(\tilde w) \,\frac{\rd^2 \tilde w}{2i\,\IM\,\ttau}}_\ttau
= \Vev{ \iint \left(\frac{\partial w}{\partial\tilde w}\right)^{h}J(w)
   \,\frac{\rd^2 w}{2i\,\IM\,\tau}}_{\tau}
= 
\Vev{\int_0^1 \tau^{h} J(x) \,\rd x}_{\tau}
=\tau^{h}\Vev{Q}_{\tau}\;.
\label{eq:Sxfmn}
\ee
Together, equations \eqref{eq:Txfmn} and \eqref{eq:Sxfmn} show that 
$\Vev{Q}_\tau$ transforms as a modular form of weight $h$ under the
generators of the M\"obius group, and hence is a modular form of
weight $h$.

It will be helpful to have the transformation law for the integral of
the current $J$ around a cylinder of circumference $R$ and radius $L$
in the case of the shift $a=0$ corresponding to a rectangular region
in the complex $w$-plane and $\Re\,{\tau}=0$.
A simple scaling gives
\be
  \frac 1 R  \Vev{ \int_0^R \rd x \,J(x) }_{R,i L}
= \frac 1 L  \Vev{ \int_0^L \rd x \,J(x) }_{L,i R}
\;.
\ee
where the subscripts denote the periods of the torus.

\subsection{Chiral CFT}

Up to now, the traces in \eqref{eq:ZLRa}, \eqref{eq:tr2} and
\eqref{eq:GGEmt} have been over the full Hilbert space of the CFT, as
have the expectation values in \eqref{eq:GGEmt} and \eqref{eq:ev1}.
While many of the physical arguments rely on the full theory, we will
be considering CFTs for which the full Hilbert space splits into
the sum of representations of a left and right chiral algebra,
\be
 \cH_{\mathrm{full}} = \oplus_{ij} m_{ij}\,\cH_i \otimes \bar{\cH}_j
\;,
\label{eq:Hsum}
\ee
and the torus partition function is 
\be
 Z(\tau,\bar\tau) = \sum_{ij} m_{ij}\,\chi_i(\tau)\,\overline{\chi_j(\tau)}
\;,
\label{eq:Zsum}
\ee
where
\be
 \chi_i(\tau) = \Tr_{\cH_i}(q^{L_0 - c/24})
\;.
\ee
Under a modular transformation, the characters $\chi_i(\tau)$
transform as a vector-valued modular form, 
\be
  \chi_i(\tau+1) = e^{2\pi i(h_i-c/24)} \chi_i(\tau)
\;,\;\;
  \chi_i(-1/\tau) = \sum_j\,S_{ij} \chi_j(\tau)
\;,
\label{eq:modinv1}
\ee
and modular invariance of the full partition function \eqref{eq:modinv0} 
follows from the multiplicities $m_{ij}$ satisfying the necessary properties. 

We are thus interested in the trace over the spaces $\cH_i$ including a charge $Q$,
\be
  \chi_i(\tau,Q) = 
\Tr_{\cH_i}( q^{L_0 - c/24}\,e^Q\,)
\;,
\label{eq:chiral sum}
\ee
and their behaviour under a modular transform. If the chiral algebra
has commuting conserved quantities $Q_i$ and $Q = \sum_i \alpha_i
Q_i$, then \ref{eq:chiral sum} is the (full) character of the
representation $\cH_i$, and $\chi_i(\tau) = \chi_i(\tau,0)$ is the
reduced character, or simply ``the character''.

If $Q$ is the integral of a local holomorphic current, then the
partition function in
\eqref{eq:GGEmt} is given by
\be
 Z(\tau,\bar\tau,Q) = 
  \sum_{ij} m_{ij} \chi_i(\tau,Q)\,\overline{\chi_j(\tau)}
\;,
\ee
and the conjectured behaviour \eqref{eq:GGEmt} corresponds to
\be
  \chi_i(\tau+1,Q) = e^{2\pi i(h_i-c/24)} \chi_i(\tau,Q)
\;,\;\;
  \chi_i(-1/\tau,Q) = \sum_j\,S_{ij} \chi_j(\tau,\tilde Q)
\;,
\ee
with the same matrices $S_{ij}$ that appear in the usual rules for the
characters, \eqref{eq:modinv1}.

\section{The Free fermion model, KdV charges and GGE}
\label{sec:2}

In this section we describe the free fermion model, find expressions
for the KdV charges and hence construct the GGE partition function.
It is easy to find expressions for the currents associated to the KdV
charges as they are quasi-primary fields bilinear in the fermion field
and this determines them uniquely, up to normalisation. 
The first problem though is to
find the action of the charges defined on the cylinder when mapped to
the plane, and we use an indirect argument to show that the KdV
charges on the cylinder have an especially simple form \eqref{eq:2.2.18}
on the plane. Since these charges are diagonal when acting on the
fermion modes, this means it is then straightforward to write down the
GGE partition functions \eqref{eq:3.1.1}.

We start by recalling the free fermion model, by which we mean the
free Majorana fermion in Euclidean space. This has two components, a
holomorphic field $\psi(z)$ and an antiholomorphic field
$\bp(\bz)$. These fields have the operator product
expansions (OPE)
\be
 \psi(z)\,\psi(w) = \frac{1}{z-w} + O(1)
\;,\;\;
 \bar\psi(\bar z)\,\bar\psi(\bar w) = \frac{1}{\bar z-\bar w} + O(1)
\;.
\label{eq:OPE}
\ee
The periodicity of the fields depends on the surface on which
they live; on the cylinder, the free-fermion can be periodic (called
the Ramond sector) or antiperiodic (the Neveu-Schwarz sector).
This periodicity is swapped on mapping to the plane, so that the 
mode expansions on the plane are
\be
\psi(z) 
  = \sum_m \psi_m z^{-m-\tfrac 12}\;,
\,\;\;
\bar\psi(\bar z) 
  = \sum_m \bar\psi_m \bar z^{-m-\tfrac 12}\;,
\label{eq:ffmodes}
\ee
where $n\in\mathbb Z$ for the R sector and $n\in\mathbb Z + \tfrac
12$ in the NS sector. 
From \eqref{eq:OPE}, these modes have 
anticommutation relations
\be
\{ \psi_m,\psi_n \} = 
\{ \bar\psi_m,\bar\psi_n \} = \delta_{m+n,0}
\;,\;\;
\{ \psi_m,\bar\psi_n \} =0.
\;.
\ee
While we will be considering partition functions defined on the
cylinder or torus, we will always map the corresponding fields to the
plane and use the mode expansions \eqref{eq:ffmodes}.
We will only ever consider charges given as integrals of holomorphic
currents, and so we would like to be able to deal exclusively with the
holomorphic field $\psi(z)$ but for technical reasons there is not a
simple chiral split in the Ramond sector since the smallest
representation of the algebra  of the zero modes $\psi_0,\bar\psi_0$
is two-dimensional. For the most part this will
not be a problem and will simply introduce extra factors of 2 or 1/2
in some formulae. For more details see appendix \ref{app:ff}.

\subsection{The KdV charges}

We will start by constructing the KdV charges for the free fermion. 
Each charge $I_{2n-1}$ will be the integral around the cylinder of a quasi-primary field $J_{2n}$,
\begin{equation}
I_{2n-1} = \int_0^{2\pi i}\frac{\rd w}{2\pi i} \, J_{2n}(w)
\;,
\label{eq:2.2.1}
\end{equation}
where for the rest of this section we will assume that the cylinder
has circumference $2\pi$.

The simplest charge is the integral of the Stress tensor itself,
\be
I_1 = \int_0^{2\pi i}\frac{\rd w}{2\pi i}\, T(w)
\;.
\ee
Under the change of coordinates,  $z = \exp(w)$, we find from \eqref{eq:Txform}
\be
I_1 = \oint (z^2 T(z) - \tfrac{c}{24}) \frac{\rd z}{2\pi i z}
    = L_0 - \tfrac c{24}
\;.
\ee

The next simplest charge is the integral of the quasi-primary field
$\Lambda(z)$,
\be
 \Lambda(w) = (T(w)T(w)) - \frac{3}{10} T''(w)
\;,
\ee
where  $(\cdots)$ denotes the standard normal ordering as defined in
\cite{yellow}. 
The result of the map $z=\exp(w)$ is given in \cite{Gaberdiel},
\be
  \Lambda(w) 
= z^4 \Lambda(z) - \tfrac{5c + 22}{60} z^2 T(z) + \tfrac{c(5c+22)}{2880}
\ee
and the integral around the cylinder is
\be
\lambda_0 
= \int_0^{2\pi i} \Lambda(w) \frac{\rd w}{2\pi i}
= \Lambda_0 - \tfrac{5c+22}{60} L_0 + \tfrac{c(5c+22)}{2880}
\;.
\ee
These modes have the nice feature that their commutation relations with the free fermion modes $\psi_m$ are simple:
\be
 [I_1,\psi_m] = - m \psi_m
\;,\;\;
 [\lambda_0,\psi_m] = - \frac 76 m^3 \psi_m
\;,
\ee
Rescaling $\lambda_0$ to $I_3 = (6/7)\lambda_0$, we have
\be
 [I_1,\psi_m] = - m \psi_m
\;,\;\;
 [I_3,\psi_m] = - m^3 \psi_m
\;,
\ee
The fact that the charges are diagonal on the fermion modes is a
consequence of the fact in the free fermion model, the fields $T(z)$ and $\Lambda(z)$ are
both bilinear in the fermion field,
\be
 T(z) = \tfrac 12 (\psi'\psi)
\;,\;\;
\tfrac 67 \Lambda(z) = \tfrac 12 (\psi'''\psi) - \tfrac
9{20}(\psi''\psi)'
\;.
\ee
This leads to the obvious identification of the KdV charges in the
free fermion model as the integrals of the following currents,
bilinear in the fermion fields:
\be
  I_{2n-1} = \int_{0}^{2\pi i} \frac{\rd w}{2\pi i} J_{2n}(w)
\;,\;\;
 J_{2n}(z) = \tfrac 12(\psi^{(2n-1)}\psi)(z) \text{ mod }\partial
\;,
\label{eq:J2ndef}
\ee
where $\partial=\partial_z$ and the total derivative terms are required for $J_{2n}$ to be a
quasi-primary field.

It should be clear from \eqref{eq:J2ndef} that the commutation
relations of the charges $I_{2n-1}$ with the fermion modes is
\be
 [I_{2n-1},\psi_m] = -m^{2n-1} \psi_m
\;,
\ee
and this means that the $I_{2n-1}$ takes the form
\begin{equation}
I_{2n-1}=\sum_{m>0}m^{2n-1}\psi_{-m}\psi_m-c_{2n-1}^{\text{NS/R}}
\;.
\label{eq:2.2.18}
\end{equation}
The the constants $c_{2n-1}^{\text{NS/R}}$ can be deduced from
zeta-function regularisation of the fermion mode sum on the cylinder,
as in \cite{Fendley} where the value on the ground state was
calculated in this way.  We shall instead deduce these constants from
the modular properties of the traces of the charges on the cylinder,
with the result \eqref{eq:2.2.4}.

Since it will be helpful to have explicit expressions for $J_{2n}(w)$,
we do this first, in the next section.

\subsection{Constructing quasi-primary fields}
\label{sec:2.1}

We want to construct a quasi-primary field from bilinears in the
fermion fields. Here we will do this in the complex plane. We will
also show that these quasi-primary fields define a pre-Lie algebra as
defined in \cite{Dijkgraaf}, we will use this fact later in section
\ref{sec:DME}.

We start by constructing the quasi-primary fields in the plane, in the
NS sector. 
We will construct the weight $2n$ quasi-primary fields $J_{2n}(z)$ from
the weight $2n$ fermion bilinears  
\begin{equation}
(\psi^{(2n-k-1)}\psi^{(k)})(z)
\;,
\label{eq:2.1.2}
\end{equation}
where the brackets denote normal ordering as defined in
\cite{yellow}. Using the state operator correspondence we see that
this field corresponds to the state 
\begin{equation}
(\psi^{(2n-k-1)}\psi^{(k)})(0)|0\rangle=(2n-k-1)!k!\psi_{-2n+k+\frac{1}{2}}\psi_{-k-\frac{1}{2}}|0\rangle 
\;.
\label{eq:2.1.3}
\end{equation}
We define the state
\begin{equation}
|\tJ_{2n}\rangle=\sum_{k=0}^{n-1}(-1)^k{2n-1\choose k}\psi_{-2n+k+\frac{1}{2}}\psi_{-k-\frac{1}{2}}|0\rangle
\;.
\label{eq:2.1.4}
\end{equation}
Using the commutation relations 
$[L_1,\psi_{-k-\frac{1}{2}}]=k\psi_{-k+\frac{1}{2}}$
and
$[L_0,\psi_{-k-\frac{1}{2}}]=(k+\frac 12)\psi_{-k-\frac{1}{2}}$
we find $L_0|\tJ_{2n}\rangle=2n |\tJ_{2n}\rangle
$ and $L_1|\tJ_{2n}\rangle=0$, i.e. the state is quasi-primary of
weight $(2n)$. 
The state operator correspondence then gives $|\tilde
J_{2n}\rangle=\tJ_{2n}(0)|0\rangle$ for the operator 
\begin{equation}
\tJ_{2n}(z)=\sum_{k=0}^{n-1}(-1)^k\frac{1}{(2n-1)!}{2n-1\choose k}^2(\psi^{(2n-k-1)}\psi^{(k)})(z)
\;.
\label{eq:2.1.5}
\end{equation}
Hence we have found the quasi-primary fields in the plane. Since being
quasi-primary is a local property of the fields, the quasi-primary
fields take the same form \eqref{eq:2.1.5}
in the R sector. 

We now want to look at the coefficients of the first and second order
poles in the OPE 
\be
\tJ_{2n}(z)\tJ_{2m}(w)
= \sum_{k>0} \frac{ [\tJ_{2n}\tJ_{2m}]_k}{(z-w)^k} + O(1)
\;.
\label{eq:JJOPE}
\ee
This will allow
us to show that the $\tJ_{2n}(z)$ form a pre-Lie algebra, see
\cite{Dijkgraaf} for the details. If the first order poles are total
derivatives then their integral can be used to construct a pre-Lie
algebra structure. The free fermion OPE is given in \eqref{eq:OPE}.
Using this we find the OPE of bilinears of the form \eqref{eq:2.1.2} is
\begin{eqnarray}
&(\psi^{(2n-k-1)}\psi^{(k)})(z)(\psi^{(2m-l-1)}\psi^{(l)})(w)=\dots
\nonumber\\
&+(-1)^k((2n{+}4m{-}2l{-}3)(\psi^{(2n+2m-l-3)}\psi^{(l)})(w){-}(2n{+}2l{-}1)(\psi^{(2n+l-2)}\psi^{(2m-l-1)})(w))\frac{1}{(z-w)^2}
\nonumber\\
&+2(-1)^k((\psi^{(2n+2m-l-2)}\psi^{(l)})(w){-}(\psi^{(2n+l-1)}\psi^{(2m-l-1)})(w))\frac{1}{z-w}+O(1)
\;,
\label{eq:2.1.7}
\end{eqnarray}
where the dots are higher order poles. This then allows us to
calculate the coefficients of the first and second order poles in 
\eqref{eq:JJOPE},
\begin{eqnarray}
\left[\tJ_{2n}\tJ_{2m}\right]_1&=&0 \text{ mod }\partial
\;,
\label{eq:J tilde 1}\\
\partial^{-1}\left[\tJ_{2n}\tilde
  J_{2m}\right]_1&=&\frac{2n-1}{2(2n-1)!(2m-1)!}{4n-2\choose
  2n-1}{4m-2\choose 2m-1}(\psi^{(2n+2m-3)}\psi)\text{ mod }\partial
\;,\label{eq:J tilde 1 integrated}
\\
\left[\tJ_{2n}\tJ_{2m}\right]_2&=&\frac{n+m-1}{(2n-1)!(2m-1)!}{4n-2\choose 2n-1}{4m-2\choose 2m-1}(\psi^{(2n+2m-3)}\psi)\text{ mod }\partial
\;,\label{eq:J tilde 2}
\end{eqnarray}
where
$\partial=\frac{\partial}{\partial z}$. 

We can now re-scale the fields $\tJ_{2n}$ to get
\begin{equation}
J_{2n}=(2n-1)!{4n-2\choose 2n-1}^{-1}\tilde
J_{2n}=\frac{1}{2}(\psi^{(2n-1)}\psi)\text{ mod }\partial
\;.
\label{eq:2.1.9}
\end{equation}
The coefficients in (\ref{eq:J tilde 1}--\ref{eq:J tilde 2}) become
\begin{eqnarray}
\left[J_{2n}J_{2m}\right]_1&=&0\text{ mod }\partial
\label{eq:2.1.10}
\;,
\\
\partial^{-1}\left[J_{2n}J_{2m}\right]_1&=&(2n-1)J_{2(n+m-1)}
\text{ mod }\partial
\;,
\label{eq:2.1.11}\\
\left[J_{2n}J_{2m}\right]_2&=&(2n+2m-2)J_{2(n+m-1)}\text{ mod }\partial
\label{eq:2.1.12}
\;.
\end{eqnarray}
Using the definition for the connection and corresponding Lie bracket given in \cite{Dijkgraaf}, if we define $\nabla_{J_{2m}}=\nabla_{2m}$ we have
\begin{eqnarray}
\nabla_{2m}J_{2n}&=&(2n-1)J_{2(n+m-1)} \text{ mod }\partial
\;,
\label{eq:2.1.13}
\\
\left[J_{2m},J_{2n}\right]&=&\nabla_{2m}J_{2n}-\nabla_{2n}J_{2m}=2(n-m)J_{2(n+m-1)}\text{
  mod }\partial 
\;.
\label{eq:2.1.14}
\end{eqnarray}
From \eqref{eq:2.1.14} we see that the space of fields generated by $J_{2n}$ is closed under the Lie bracket. The $J_{2n}$ generate the pre-Lie algebra $W$ as defined in \cite{Dijkgraaf}. Here the pre-Lie algebra is isomorphic to the space of odd holomorphic vector fields with the identification
\begin{equation}
J_{2n}\sim z^{2n-1}\frac{\partial}{\partial z}
\;.
\label{eq:2.1.15}
\end{equation}
We will make use of this isomorphism in section \ref{sec:DME}.

\subsection{Constructing KdV charges}
\label{aec:2.2}

Having found expressions for the quasi-primary field $J_{2n}(w)$, 
we now want to express the charge $I_{2n-1}$ defined by the integral
around the cylinder 
\be
    I_{2n-1} = \int_{0}^{2\pi i} \frac{\rd w}{2\pi i} J_{2n}(w)
\;,
\ee
in terms of the modes of the fermion field on the plane.
The current $J_{2n}(w)$ is quasi-primary but not primary,
and so we need again the transformation formula for conformal fields
given in \cite{Gaberdiel}. If $z=e^w$ with $w$ the coordinate on the
torus, then $J_{2n}(w)$ is 
\begin{equation}
\left.  J_{2n}(w) \right|_{\text{cylinder}}
= \frac{1}{2}z^{2n}(\psi^{(2n-1)}\psi)(z)+\sum_{k=1}^{n-1}z^{2k}\sum_{l=0}^{k-1}\alpha_{k,l}(\psi^{(2k-1-l)}\psi^{(l)})(z)+\alpha_0\text{ mod }z\partial
\;,
\label{eq:2.2.5}
\end{equation}
for some constants $\alpha_{k,l}$ and $\alpha_0$, where mod $z\partial$ means we drop terms of the form $z\frac{d}{dz}(\dots)$.
One can see that these are constants from the recursive definition of
the functions $T_n(z)$ in \cite{Gaberdiel} and 
$T_1(z) =
\frac{f''(z)}{2f'(z)}=\frac{1}{2}$ for $f(z)=e^z$.
For details, see
\cite{Gaberdiel}.
There are no terms
of the form $z^{2k+1}(\psi^{(2k-l)}\psi^{(l)})$ in \eqref{eq:2.2.5}
since modulo total derivatives we can replace them with terms of the
form $z^{2k}(\psi^{(2k-1-l)}\psi^{(l)})$. Since $z=e^w$, the integral
over $w$ can be written as an integral over $z$
\begin{equation}
 \int_0^{2\pi i}\frac{\rd w}{2\pi i}J_{2n}(w)
=\oint\frac{\rd z}{2\pi iz}J_{2n}(w)
\;,
\label{eq:2.2.6}
\end{equation}
where we use the form of $J_{2n}(w)$ in terms of $z$ from
\eqref{eq:2.2.5}. Any terms of $\frac{1}{z}J_{2n}(w)$ that are a total
derivative with respect to $z$ vanish when integrated. The zero mode
of $(\psi^{(2k-1-l)}\psi^{(l)})$ is 
\begin{equation}
 (\psi^{(2k-1-l)}\psi^{(l)})_0
=(-1)^l\sum_{m\geq0}\left(2m^{2k-1}+(\text{odd
    powers of }m\text{ of order}<2k{-}1)\,\right) \psi_{-m}\psi_m
\;.
\label{eq:2.2.7}
\end{equation}
Hence the charge $I_{2n-1}$
is of the form
\begin{equation}
I_{2n-1}
=\sum_{m\geq0}m^{2n-1}\psi_{-m}\psi_m+\sum_{k=1}^{n-1}\beta_k\sum_{m\geq0}m^{2k-1}\psi_{-m}\psi_m+\beta_0
\;,
\label{eq:2.2.8}
\end{equation}
where the $\beta_k$ can be related to the $\alpha_{k,l}$ and
$\alpha_0$ from \eqref{eq:2.2.5}.
\\

From \cite{Ross} and \cite{Dijkgraaf} we have that the expectation
value on the torus
\be
\left\langle \int_0^{2\pi i}\frac{\rd w}{2\pi i}J_{2n}(w)
\right\rangle_{\tau}
\ee
is a modular form of weight $2n$ (for an introduction to modular
forms see appendix \ref{app:mf}). 

We will want to sum over 
the two sectors, with and
without the insertion of $(-1)^F$,
and correspondingly define the traces $\Tr_{\NS/\R,\pm}$ in equations
\eqref{eq:NSTrdef} and \eqref{eq:RTrdef}. 
It will also be helpful
to factor off the ``fermion character'' so we 
introduce
the expectation value of an operator $\mathcal{O}$
as
\begin{equation}
\langle\mathcal{O}\rangle^{\NS/\R,\pm}(\tau)
= 
\begin{cases}
 \frac{\Tr_{\NS/\R,\pm}(\,\cO\,q^{L_0-c/24}) }{\chi^{\NS/\R,\pm}(\tau)}
\;, & \hbox{not $\R,-$}  \\
0 & \R,-
\end{cases}
\label{eq:3.1.3}
\end{equation}
where $\chi^{\text{NS/R},\pm}(\tau)$ are the fermion characters
defined in \eqref{eq:NS+def}, \eqref{eq:NS-def} and \eqref{eq:R+def}, \eqref{eq:R-def}. 
The argument $\tau$ on the left hand
side is to indicate that $q=e^{2\pi i\tau}$; when the value of $\tau$
is clear from context, we will omit it. 
Following the discussion in appendix \ref{app:ff},
if $I_{2n-1}$ is composed entirely of modes of the holomorphic field
$\psi(z)$, we can calculate its expectation value, $\langle I_{2n-1} \rangle_\tau$, as 
\be
Z(\tau,\bar\tau)\langle I_{2n-1} \rangle_\tau
=
\tfrac 12\, |{\chi^{\NS,+}}|^2\, \langle I_{2n-1}\rangle^{\text{NS},+}
+ 
\tfrac 12\,|{\chi^{\NS,-}}|^2\, \langle I_{2n-1}\rangle^{\text{NS},-}
+ 
\tfrac 14\, |{\chi^{\R,+}}|^2\, \langle I_{2n-1}\rangle^{\text{R},+}\;.
\label{eq:Isum}
\ee
The characters $\chi^{\NS/\R,\pm}$ are a vector-valued modular form of
weight 0 under the full modular group $\Gamma_1$=SL$(2,\mathbb{Z})$,
but they are also modular forms of weight 0 under the congruence subgroup 
$\Gamma(2)$ (for definitions of modular forms and $\Gamma(2)$ see appendix \ref{app:mf}).
The whole expression \eqref{eq:Isum} being an element of
$M_{2n}(\Gamma_1)$ (the space of  modular forms of weight
$2n$ under $\Gamma_1$) means that the traces 
\be
  \Tr_{\NS,\pm}(I_{2n-1}q^{L_0-c/24})
\;,\;\;\;\;
  \Tr_{\R,+}(I_{2n-1}q^{L_0-c/24})
\;,
\ee
must be a vector-valued modular form of weight $2n$ under $\Gamma_1$, 
and each must also be an element of $M_{2n}(\Gamma(2))$ (the space of
modular forms of weight $2n$ under $\Gamma(2)$). 
The results are the same, whichever result we enforce.

Using the result 
\be
 \Tr_{\NS/\R,\pm}( m^k \psi_{-k}\psi_k\,q^{L_0-c/24})
= \pm\frac{m^k q^k}{1 \pm q^k} \chi^{\NS/R,\pm}
\;,
\ee
and the expression 
\eqref{eq:2.2.8},
we find
\be
\Tr_{\NS/R,\pm}(I_{2n-1}\,q^{L_0-c/24})
=
\chi^{\text{NS/R},\pm}(\tau)\left(\beta_0\pm \sum_{m\geq0}\frac{m^{2n-1}q^m}{1\pm q^m}\pm\sum_{k=1}^{n-1}\beta_k\sum_{m\geq0}\frac{m^{2k-1}q^m}{1\pm q^m}\right)
\label{eq:tr12}
\ee

If we define
\begin{equation}
c_{2n-1}^{\NS}=\tfrac{1}{2}\left(\left(\tfrac{1}{2}\right)^{2n-1}{-}1\right)\zeta(1-2n)
\;,\;\;\;\;
c_{2n-1}^{\R}=\tfrac{1}{2}\zeta(1-2n)
\;,
\label{eq:2.2.4}
\end{equation}
where $\zeta(z)$ is the Riemann zeta function, then
we 
can rewrite the series in \eqref{eq:tr12}
in
terms of Eisenstein series (see appendix \ref{app:mf} for definitions) as follows 
\begin{eqnarray}
\sum_{m=0}\frac{m^{2n-1}q^m}{1+q^m}-c_{2n-1}^{\text{R}}\!\!\!\!\!&=&\!\!\!\!\!\frac{\zeta(1-2n)}{2}(E_{2n}(\tau)-2E_{2n}(2\tau))
\;,
\label{eq:2.2.12}
\\
-\!\!\sum_{m=0}\frac{(m{+}\frac{1}{2})^{2n-1}q^{m+\frac{1}{2}}}{1{-}q^{m+\frac{1}{2}}}{-}c_{2n-1}^{\text{NS}}\!\!\!\!\!&=&\!\!\!\!\!\frac{\zeta(1-2n)}{2^{2n}}\left(2^{2n-1}E_{2n}(\tau)-E_{2n}\left(\frac{1}{2}\tau\right)\right)
\;,
\label{eq:2.2.13}
\\ 
\sum_{m=0}\frac{(m{+}\frac{1}{2})^{2n{-}1}q^{m+\frac{1}{2}}}{1{+}q^{m+\frac{1}{2}}}{-}c_{2n-1}^{\text{NS}}\!\!\!\!\!&=&\!\!\!\!\!\frac{\zeta(1{-}2n)}{2^{2n}}\!\left(\!E_{2n}\!\left(\frac{\tau}{2}\right){+}2^{2n}E_{2n}(2\tau){-}(2^{2n-1}{+}2)E_{2n}(\tau)\!\right)\!\!\!\;.
\label{eq:2.2.14}
\end{eqnarray}
Each of the above combinations of Eisenstein series is a modular form
of weight $2n$ under the group $\Gamma(2)$. 
If we denote these combinations of Eisenstein series by
$F^{\text{NS/R},\pm}_{2n}(\tau)$, 
\begin{equation}
  F^{\text{NS/R},\pm}_{2n}(\tau)
= \pm\sum_{m\geq0}\frac{m^{2n-1}q^m}{1\pm
  q^m}-c_{2n-1}^{\text{NS/R}}
\;,
\label{eq:2.2.15}
\end{equation}
where for NS we sum over half integers ($m\in\mathbb{Z}+\frac{1}{2}$) and for R we sum over integers ($m\in\mathbb{Z}$), then we have
\begin{equation}
\text{Tr}_{\text{NS/R},\pm}(I_{2n-1} q^{L_0 -c/24})
=\chi^{\text{NS/R},\pm}(\tau)\!\left(\!\beta_0{+}(F^{\text{NS/R},\pm}_{2n}{+}c_{2n-1}^\text{NS/R}){+}\sum_{k=1}^{n-1}\beta_k(F^{\text{NS/R},\pm}_{2k}{+}c_{2k-1}^{\text{NS/R}})\!\right)\!
\;.
\label{eq:2.2.16}
\end{equation}
We can see that this is a modular form of weight $2n$ for
$\Gamma(2)$ if and only if
\be
\beta_0=-c_{2n-1}^{\text{NS/R}}
\;,\;\;\;\;
\beta_k=0 \text{ for } k>0
\;,
\label{eq:2.2.17}
\end{equation}
Hence we have proven that the charges $I_{2n-1}$ take the form
\begin{equation}
I_{2n-1}=\sum_{m>0}m^{2n-1}\psi_{-m}\psi_m-c_{2n-1}^{\text{NS/R}}
\;.
\label{eq:2.2.18b}
\end{equation}
where the constants are given by \eqref{eq:2.2.4}.
The values of these charges on the ground states agree with the
results in \cite{Fendley}
obtained by zeta function regularisation. 

Having obtained the charges, we can now find the GGE averages easily,
as the charges are diagonal on the fermion modes.

\section{The GGE average and its modular transform}
\label{sec:MT1}
Now that we have the charges we would like to investigate the modular
properties of a GGE which has one of these charges inserted. We will
find that an asymptotic expression for the modular transformed GGE can
be obtained.
Initially we will just look at transforms under the group $\Gamma(2)$ so the different sectors don't map into each other. At the end of this section we will see that under the full modular group $\Gamma_1$ the different sectors map into each other in the same way they do for free fermions.

We want to consider how the presence of the KdV charges $I_{2n-1}$ in the free
fermion partition function will affect the modular properties of these
partition functions. 
We can find the asymptotic expansion of this GGE in the
chemical potential of the KdV charges and we see that the coefficients
are quasi-modular forms whose modular transform we can
determine. These coefficients are given in terms of complete
exponential Bell polynomials so using the generating function for
these polynomials allows us to find a formal expression for the
modular transform of the GGE. However the series that appear in this
expression are divergent but we find expressions in terms of
hypergeometric functions whose asymptotic series are those that appear
in the GGE. When we consider the case of a rectangular torus ($\tau$ is pure imaginary) we find that the exact modular transformed GGE is real but the expression in terms of hypergeometric functions is complex so we can immediately see that they don't match. However there have already been investigations into these KdV
charges in \cite{Ross} and \cite{Dijkgraaf} and we show in section \ref{sec:Dijk} that our asymptotic
results match what has previously been found. 

\subsection{Explicit expressions for the GGE averages}

The charges as given in  \eqref{eq:2.2.18} are diagonal on the fermion
modes, and so in each sector the  GGE average is simply
\begin{equation}
 \Tr_{\text{NS/R},\pm}\left(e^{\sum_{n=1} \alpha_{2n-1}I_{2n-1}}\right)
= e^{-\sum_{n=1} \alpha_{2n-1}c_{2n-1}^{\text{NS/R}}}
\, \prod_{k\geq 0}\left(1\pm e^{\sum_{n=1}\alpha_{2n-1}k^{2n-1}}\right)
\;.
\label{eq:3.1.1}
\end{equation}
(Note that this is entirely analogous to the calculation of $W_{1+\infty}$
characters in \cite{awata}.)
If we note that $I_1 = L_0 - c/24$, this can
equally be written as
\begin{equation}
 \Tr_{\text{NS/R},\pm}\left(e^{\sum_{n>1} \alpha_{2n-1}I_{2n-1}}\,q^{L_0-c/24}\right)
= e^{-\sum_{n>1} \alpha_{2n-1}c_{2n-1}^{\text{NS/R}}}\,q^{h-c/24}
\, \prod_{k\geq 0}\left(1\pm e^{\sum_{n>1}\alpha_{2n-1}k^{2n-1}}q^k\right)
\;,
\label{eq:3.1.1}
\end{equation}
where $q=e^{2\pi i\tau}$.

Rather than consider the general charge
\be
 Q = \sum_{n>1} \alpha_{2n-1} I_{2n-1}
\;,
\ee
we will consider the simpler case where just one of the $\alpha_{2m-1}$ is
non-zero.
Setting $\alpha_{2m-1}=\alpha$ and all other $\alpha_{2n-1}$ to zero, we get
\begin{equation}
 \Tr_{\NS/\R,\pm}(e^{\alpha I_{2m-1}}q^{I_1})
=e^{-\alpha c_{2m-1}^{\NS/\R}}q^{-c_1^{\NS/\R}}\prod_{k\geq0}(1\pm e^{\alpha k^{2m-1}}q^k)
\;.
\label{eq:3.1.2}
\end{equation}
If we put $e^\alpha = w$, this takes the nice form
\begin{equation}
 \Tr_{\NS/\R,\pm}(w^{I_{2m-1}}q^{I_1})
=w^{-c_{2m-1}^{\NS/\R}}q^{-c_1^{\NS/\R}}\prod_{k\geq0}(1\pm w^{k^{2m-1}}q^k)
\;.
\label{eq:3.1.2b}
\end{equation}
We  will always assume that the modular parameter $\tau$
has a positive imaginary part so that $|q|<1$, and so the infinite
product in \eqref{eq:3.1.2} is convergent for
$\text{Re}(\alpha)\leq0$, or equivalently $|w|\leq 1$.

\subsection{Explicit expressions for $\langle I_{2m-1}^n\rangle^{\text{NS/R},\pm}$ and
  their modular transform}
\label{sec:3.1}

As a first step to calculating the modular transform of the full GGE
average, we consider the traces of powers of the charge $I_{2m-1}$ and
find their modular transforms explicitly. 

By differentiating
\eqref{eq:3.1.2} $n+1$ times with respect to $\alpha$ and then setting
$\alpha=0$ we have the recursion relation 
\begin{equation}
\langle I_{2m-1}^{n+1}\rangle^{\text{NS/R},\pm} = \sum_{k=0}^n{n\choose k}\langle I_{2m-1}^{n-k}\rangle^{\text{NS/R},\pm}\frac{1}{(2\pi i)^k}\frac{d^k}{d\tau^k}F^{\text{NS/R},\pm}_{2(km+m-k)}
\;,
\label{eq:3.1.4}
\end{equation}
where $F^{\text{NS/R},\pm}_{2k}$ is as defined in \eqref{eq:2.2.15}. Using \eqref{eq:3.1.4} with $n=0$ we can see that
\begin{equation}
F_{2m}^{\text{NS/R},\pm} = \langle I_{2m-1}\rangle^{\text{NS/R},\pm}
\;.
\label{eq:3.1.5}
\end{equation}
Hence we can eliminate the $F_{2m}^{\text{NS/R},\pm}$ in favour of $\langle I_{2m-1}\rangle^{\text{NS/R},\pm}$ to get the recursion relation
\begin{equation}
\langle I_{2m-1}^{n+1}\rangle^{\text{NS/R},\pm} = \sum_{k=0}^n{n\choose k}\langle I_{2m-1}^{n-k}\rangle^{\text{NS/R},\pm}\frac{1}{(2\pi i)^k}\frac{d^k}{d\tau^k}\langle I_{2(km+m-k)-1}\rangle^{\text{NS/R},\pm}
\;.
\label{eq:3.1.6}
\end{equation}
The recursion relation \eqref{eq:3.1.6} is the same as the recursion relation satisfied by the complete exponential Bell polynomials. The complete exponential Bell polynomials, $B_n(x_1,\dots,x_n)$, are defined in \cite{Bell}. Their generating function is
\begin{equation}
\exp\left(\sum_{n=1}^\infty \left(x_n\frac{\alpha^n}{n!}\right)\right) = \sum_{n=0}^\infty B_n(x_1,\dots,x_n)\frac{\alpha^n}{n!}
\;,
\label{eq:3.1.7}
\end{equation}
and they satisfy the recursion relation
\begin{equation}
B_{n+1}(x_1,\dots,x_{n+1})=\sum_{k=0}^n{n\choose k}B_{n-k}(x_1,\dots,x_{n-k})x_{k+1}
\;.
\label{eq:3.1.8}
\end{equation}
Using recursion relation \eqref{eq:3.1.6} and the fact $\langle1\rangle^{\text{NS/R},\pm}=1$ from the definition \eqref{eq:3.1.3}, we see that
\begin{eqnarray}
&\langle I_{2m-1}^n\rangle^{\text{NS/R},\pm}= \nonumber\\
&B_n \left(\langle
I_{2m-1}\rangle^{\text{NS/R},\pm},\frac{1}{2\pi i}\frac{d}{d\tau}\langle
I_{4m-3}\rangle^{\text{NS/R},\pm},\dots,\frac{1}{(2\pi
  i)^{n-1}}\frac{d^{n-1}}{d\tau^{n-1}}\langle
I_{2n(m-1)+1}\rangle^{\text{NS/R},\pm}\right)
\;. 
\label{eq:3.1.9}
\end{eqnarray}
As mentioned above, the infinite product in \eqref{eq:3.1.2} only
converges for $\Re(\alpha)\leq0$. If we expand \eqref{eq:3.1.2} as a
power series in $\alpha$, the series must therefore have a zero radius of
convergence. We therefore have an asymptotic expansion for $\langle
e^{\alpha I_{2m-1}}\rangle^{\text{NS/R},\pm}$ about $\alpha=0$ 
\begin{equation}
     \langle e^{\alpha I_{2m-1}}\rangle^{\text{NS/R},\pm}
\sim \sum_{n=0}^\infty\frac{\alpha^n}{n!}\langle I_{2m-1}^n\rangle^{\text{NS/R},\pm}
\;.
\label{eq:3.1.10}
\end{equation}
Hence, using the generating function \eqref{eq:3.1.7}, we have the
formal expression for the expectation value of 
$\langle e^{\alpha   I_{2m-1}}\rangle^{\text{NS/R},\pm}$ in terms of $\tau$ derivatives
of the expectation values $\langle I_{2n(m-1)+1}\rangle^{\text{NS/R},\pm}$ 
\begin{equation}
\langle e^{\alpha I_{2m-1}}\rangle^{\text{NS/R},\pm} \sim \exp\left(\sum_{n=1}^\infty\left(\frac{\alpha^n}{n!(2\pi i)^{n-1}}\frac{d^{n-1}}{d\tau^{n-1}}\langle I_{2n(m-1)+1}\rangle^{\text{NS/R},\pm}\right)\right)
\;.
\label{eq:3.1.11}
\end{equation}
Since $F_{2m}^{\text{NS/R},\pm}=\langle I_{2m-1}\rangle^{\text{NS/R},\pm}$ we have that $\langle I_{2m-1}\rangle^{\text{NS/R},\pm}\in M_{2m}(\Gamma(2))$. We want the modular transform of $\frac{d^n}{d\hat\tau^n}\langle I_{2m-1}\rangle^{\text{NS/R},\pm}(\hat\tau)$ which is given in \cite{Zagier}. If we consider the modular transform $\hat\tau=\frac{a\tau+b}{c\tau+d}$ where $\big(\begin{smallmatrix}
  a & b\\
  c & d
\end{smallmatrix}\big)\in\Gamma(2)$ then we find
\begin{equation}
\frac{d^n}{d\hat\tau^n}\langle I_{2m-1}\rangle^{\text{NS/R},\pm}(\hat\tau)=\sum_{r=0}^n{n\choose r}\frac{(2m{+}n{-}1)!}{(2m{+}r{-}1)!}c^{n-r}(c\tau{+}d)^{2m+n+r}\frac{d^r}{d\tau^r}\langle I_{2m-1}\rangle^{\text{NS/R},\pm}(\tau)
\;.
\label{eq:3.1.12}
\end{equation}
Hence we have the modular transform of \eqref{eq:3.1.11}
\begin{eqnarray}
&\langle e^{\alpha
  I_{2m-1}}\rangle^{\text{NS/R},\pm}(\hat\tau)\sim\label{eq:3.1.3}\\
&\exp\!\left(\sum_{n=1}^\infty\frac{\alpha^n}{n!(2\pi 
    i)^{n{-}1}}\sum_{r=0}^{n-1}{n{-}1\choose
    r}\frac{(n(2m{-}1))!c^{n-1-r}}{(2n(m-1){+}r{+}1)!}(c\tau{+}d)^{n(2m-1)+r+1}\frac{d^r}{d\tau^r}\langle 
  I_{2n(m-1)+1}\rangle^{\text{NS/R},\pm}\!\right)\!\nonumber\;,
\end{eqnarray}
which is again a formal expression. We can instead consider the
asymptotic series of $\langle e^{\alpha I_{2m-1}}\rangle^{\text{NS/R},\pm}(\hat\tau)$
about $\alpha=0$ from \eqref{eq:3.1.10} where the coefficients are
given by \eqref{eq:3.1.9}. Using \eqref{eq:3.1.12} we can rewrite the
coefficients in terms of $\frac{d^k}{d\tau^k}\langle
I_{2n-1}\rangle^{\text{NS/R},\pm}(\tau)$ so we have an asymptotic series for $\langle
e^{\alpha I_{2m-1}}\rangle^{\text{NS/R},\pm}(\hat\tau)$ about $\alpha=0$ where the
coefficients are given in terms of $\frac{d^k}{d\tau^k}\langle
I_{2n-1}\rangle^{\text{NS/R},\pm}(\tau)$. 

\subsection{The modular transform of the GGE average}
\label{sec:3.2} 
 
We want to write the modular transform of $\langle e^{\alpha
  I_{2m-1}}\rangle^{\text{NS/R},\pm}$, under the modular transform $\gamma\in\Gamma(2)$
in the form 
\begin{equation}
\langle e^{\alpha I_{2m-1}}\rangle^{\text{NS/R},\pm}(\hat\tau)=\langle e^{Q_m(\alpha,\tau,\gamma)}\rangle^{\text{NS/R},\pm}(\tau)
\;,
\label{eq:3.2.1}
\end{equation}
for some operator $Q_m$ that we need to determine. We will assume that the $Q_m$ can be expanded as an asymptotic series in the $I_{2m-1}$ charges. We will provide evidence for this assumption when we look at Dijkgraaf's master equation in section \ref{sec:DME}. As an asymptotic expansion we have
\begin{equation}
Q_m(\alpha,\tau,\gamma)\sim\sum_{n=1}\alpha^{(m)}_{2n+1}I_{2n+1}
\;,
\label{eq:3.2.2}
\end{equation} 
where the coefficients $\alpha^{(m)}_{2n+1}$ depend on $\alpha$,
$\tau$ and the elements in $\gamma$. We can write \eqref{eq:3.2.1} as an asymptotic series in powers of the operator $Q_m$
\begin{equation}
\langle e^{\alpha I_{2m-1}}\rangle^{\text{NS/R},\pm}(\hat\tau)\sim\sum_{n=0}^\infty\frac{1}{n!}\langle Q_m^n\rangle^{\text{NS/R},\pm}(\tau)\;,
\label{eq:Qasym}
\end{equation}
and we will try to match this to the asymptotic series \eqref{eq:3.1.10}. We can substitute \eqref{eq:3.2.2} into \eqref{eq:Qasym} to get an asymptotic series about $\alpha=0$ for $\langle e^{\alpha I_{2m-1}}\rangle^{\text{NS/R},\pm}(\hat\tau)$ in terms of expectation values $\langle I_{n_1}\dots I_{n_I}\rangle^{\text{NS/R},\pm}(\tau)$. By matching the coefficients of the $\langle I_{2n-1}\rangle^{\text{NS/R},\pm}$ expectation values in the expansions \eqref{eq:3.1.10} and \eqref{eq:Qasym} we will fix the values of the $\alpha^{(m)}_{2n+1}$. Using the expression for $\langle
I_{2m-1}^n\rangle^{\text{NS/R},\pm}$ in terms of the Bell polynomials, 
\eqref{eq:3.1.9}, and the modular transform of $\frac{d^{n-1}}{d\tau^{n-1}}\langle I_{2n(m-1)+1}\rangle^{\text{NS/R},\pm}$ given in \eqref{eq:3.1.12} we can see that one of the terms in $\langle I_{2m-1}^n\rangle^{\text{NS/R},\pm}(\hat\tau)$ is going to be
\begin{equation}
\frac{(n(2m-1))!}{(2n(m-1)+1)!}\left(\frac{c}{2\pi
  i}\right)^{n-1}(c\tau+d)^{n(2m-1)+1}\langle
I_{2n(m-1)+1}\rangle^{\text{NS/R},\pm}(\tau) 
\;.
\label{eq:3.2.3}
\end{equation}
Forcing these terms to match in the asymptotic expansion
\eqref{eq:Qasym} fixes the coefficients $\alpha^{(m)}_{2p+1}$ to take
the form 
\begin{equation}
\alpha_{2p+1}=
\begin{cases}
\;\;\frac{\alpha^n}{n!(2\pi
  i)^{n-1}}\frac{(n(2m-1))!c^{n-1}}{(2n(m-1)+1)!}(c\tau+d)^{n(2m-1)+1}
\;,\;\;& p=n(m-1)
\;,\\
\;\;0 & \text{otherwise.}
\end{cases}
\label{eq:3.2.4}
\end{equation}

When we take the trace over the different fermion sectors we have the explicit formula for the modular transform of the exponential
\begin{eqnarray}
&\langle e^{\alpha I_{2m-1}}\rangle^{\text{NS/R},\pm}(\hat\tau)\sim\nonumber\\
&\frac{1}{\chi^{\text{NS/R},\pm}(\tau)}e^{-\sum_{n=1}\alpha^{(m)}_{2n(m-1)+1}c_{2n(m-1)+1}^{\text{NS/R}}}q^{-c_1^\text{NS/R}}\prod_{k\geq 0}\!\left(\!\!1\pm e^{\sum_{n=1}\alpha^{(m)}_{2n(m-1)+1}k^{2n(m-1)+1}}q^k\!\!\right)
\;\!\!.
\label{eq:3.2.5}
\end{eqnarray}
Here we have used the fact that in each of the fermion sectors we have $\chi^{\text{NS/R},\pm}(\hat\tau)=\chi^{\text{NS/R},\pm}(\tau)$ since we are considering modular transforms in the group $\Gamma(2)$. We can see that there are convergence problems with the above expression, this is expected since our results match as asymptotic series with zero radius of convergence. First, let us look at the series
\begin{equation}
\sum_{n=1}\alpha^{(m)}_{2n(m-1)+1}k^{2n(m-1)+1}
\;.
\label{eq:3.2.6}
\end{equation}
For fixed $k$ the series converges if
\begin{equation}
|\alpha|<\frac{(2m-2)^{2m-2}}{k^{2m-2}(2m-1)^{2m-1}}\frac{1}{|c(c\tau+d)|^{2m-1}}
\;.
\label{eq:3.2.7}
\end{equation}
Hence as $k\rightarrow\infty$ the radius of convergence goes to zero. This means the infinite product in \eqref{eq:3.2.5} is only defined for $\alpha=0$. Now consider the series
\begin{equation}
\sum_{n=1}\alpha^{(m)}_{2n(m-1)+1}c_{2n(m-1)+1}^{\text{NS/R}}
\;.
\label{eq:3.2.8}
\end{equation}
The asymptotic behaviour of $\zeta(1-2n)$ as $n\rightarrow\infty$ is
\begin{equation}
|\zeta(1-2n)|\sim2\sqrt{\frac{\pi}{n}}\left(\frac{n}{\pi e}\right)^{2n}
\;.
\label{eq:3.2.9}
\end{equation}
This then gives us the asymptotic behaviour of the terms in the series
\begin{equation}
|\alpha_{2n(m-1)+1}c_{2n(m-1)+1}^{\text{NS/R}}|\sim\sqrt{2m{-}1}(m{-}1)\frac{|c\tau{+}d|}{c\pi e}\!\left(\!\frac{\alpha c|c\tau{+}d|^{2m-1}(2m{-}1)^{2m-1}n^{2m-2}}{(2\pi e)^{2m-2}}\!\right)^n
\;.
\label{eq:3.2.10}
\end{equation}
As $n\rightarrow\infty$ the terms in the series diverge for all values of $m$ and $\alpha$. Again we have a series with a zero radius of convergence.

We will now show that the series in \eqref{eq:3.2.6} and \eqref{eq:3.2.8} can be expressed in terms of generalised hypergeometric functions
\begin{equation}
{}_pF_q(a_1,\dots,a_p;b_1,\dots,b_q;z)=\sum_{n=0}^\infty\frac{(a_1)_n\dots(a_p)_n}{(b_1)_n\dots(b_q)_n}\frac{z^n}{n!}
\;,
\label{eq:3.2.11}
\end{equation}
where $(a)_n$ is the Pochhammer symbol
\begin{eqnarray}
(a)_0&=&1 \;,
\nonumber\\
(a)_n&=&a(a+1)\dots(a+n-1)
\;.
\label{eq:3.2.12}
\end{eqnarray}
This will give us a finite expression whose asymptotic expansion matches \eqref{eq:3.2.5}. If $m$ is a positive integer then
\begin{eqnarray}
\left(\frac{1}{m}\right)_n\left(\frac{2}{m}\right)_n\dots\left(\frac{m-1}{m}\right)_n&=&\frac{(nm)!}{n!m^{nm}}
\;,
\label{eq:3.2.13}
\\
\left(\frac{2}{m-1}\right)_n\dots\left(\frac{m-2}{m-1}\right)_n\left(\frac{m+1}{m-1}\right)_n&=&\frac{(n(m-1)+1)!}{n!(m-1)^{n(m-1)}}
\;,
\label{eq:3.2.14}
\end{eqnarray}
and using this we define the function $f_m(z)$ as follows,
\begin{eqnarray}
&f_m(z)=\sum_{n=0}^\infty\frac{(n(2m-1))!}{n!(n(2m-2)+1)!}\left(\frac{z}{2\pi i}\right)^n\nonumber\\
&={}_{2m-2}F_{2m-3}\left(\frac{1}{2m{-}1},\dots,\frac{2m{-}2}{2m{-}1};\frac{2}{2m{-}2},\dots,\frac{2m{-}3}{2m{-}2},\frac{2m{-}1}{2m{-}2};\frac{(2m{-}1)^{2m-1}z}{2\pi i(2m{-}2)^{2m-2}}\right)
\;.
\label{eq:3.2.15}
\end{eqnarray}
Note that in the case $m=2$, $f_2(z)$ is
\begin{equation}
f_2(z)={}_2F_1\left(\frac{1}{3},\frac{2}{3};\frac{3}{2};\frac{27}{8\pi i}z\right)
\;.
\label{eq:3.2.16}
\end{equation}
We will replace the series $\sum_{n=1}\alpha^{(m)}_{2n(m-1)+1}k^{2n(m-1)+1}$ in \eqref{eq:3.2.5} with
\begin{equation}
\frac{2\pi i(c\tau+d)k}{c}\left(f_m\left(\alpha c(c\tau+d)^{2m-1}k^{2(m-1)}\right)-1\right)
\;,
\label{eq:3.2.17}
\end{equation} 
since this is the unique analytic continuation of the series.

We can rewrite the series $\sum_{n=1}\alpha^{(m)}_{2n(m-1)+1}c_{2n(m-1)+1}^{\text{NS/R}}$ in a similar fashion.
We have the following integral representations of the Riemann zeta function,
\begin{eqnarray}
  c_{2n-1}^{\NS}
  \!\!&=&\!\!\tfrac{1}{2}(2^{1-2n}{-}1)\zeta(1-2n)
  \;=\; 
 -\frac{(-1)^n}{(2\pi)^{2n}}
     \int_0^\infty\frac{t^{2n-1}}{e^t+1}\, \rd t\;,
\label{eq:3.2.18}
\\
c_{2n-1}^{\R}
  \!\!&=&\!\!
\tfrac{1}{2}\zeta(1-2n)
\;=\;
 \frac{(-1)^n}{(2\pi)^{2n}}
     \int_0^\infty\frac{t^{2n-1}}{e^t-1}\, \rd t
\;.
\label{eq:3.2.19}
\end{eqnarray}
These can be used to show that the following integrals have asymptotic expansions in $\alpha$ that match the divergent series in \eqref{eq:3.2.8}
\begin{eqnarray}
&\frac{c\tau{+}d}{2\pi ic}\int_0^\infty\frac{t}{e^t{-}1}\left(f_m\left(\frac{\alpha c(c\tau{+}d)^{2m-1}t^{2(m-1)}}{(2\pi i)^{2m-2}}\right)-1\right)dt\sim \sum_{n=1}\alpha^{(m)}_{2n(m-1)+1}c_{2n(m-1)+1}^\text{R}\;,\label{eq:Rgnd}\\
&-\frac{c\tau{+}d}{2\pi
  ic}\int_0^\infty\frac{t}{e^t{+}1}\left(f_m\left(\frac{\alpha
  c(c\tau{+}d)^{2m-1}t^{2(m-1)}}{(2\pi i)^{2m-1}}\right)-1\right)dt\sim
\sum_{n=1}\alpha^{(m)}_{2n(m-1)+1}c_{2n(m-1)+1}^\text{NS}
\;.
\label{eq:NSgnd}
\end{eqnarray}
Replacing the series in \eqref{eq:3.2.5} with the generalised hypergeometric
functions gives us a finite expression which has an asymptotic series
about $\alpha=0$ matching $\langle e^{\alpha
  I_{2m-1}}\rangle^{\text{NS/R},\pm}(\hat\tau)$.

However we can immediately see that these expressions don't match the exact modular transform of the GGE. The problem is that the expressions in terms of hypergeometric
functions give us complex values for the one particle energies. Set $m=2$ and consider a rectangular torus with purely
imaginary modular parameter, $\tau=i\tau_2$. We will consider the
modular transform $\hat\tau=\frac{-1}{\tau}$. While this is not in the
group $\Gamma(2)$ we will see in section \ref{sec:3.3} that under this
modular transformation the different sectors are transformed into each
other in the same way as the sectors are transformed into each other
when we look at the characters for the free fermions without the
charges as is done in appendix \ref{app:ff}. For this choice of the modular parameter and modular
transform the argument of the hypergeometric function is positive for
negative $\alpha$. The hypergeometric functions have a branch point
when the argument is 1 and hence as $k$ increases we pass through this
branch point and the one particle energies become complex so the asymptotic expression for the GGE becomes complex. However
the exact expression for the $S$ modular transform of the GGE,
\begin{eqnarray}
&\langle e^{\alpha I_3}\rangle(-1/\tau) = \frac{1}{\chi^{\text{NS/R},\pm}(-1/\tau)}\text{Tr}_{\text{NS/R},\pm}\left(e^{\alpha I_3}\hat{q}^{I_1}\right)=\nonumber\\
&\frac{1}{\chi^{\text{NS/R},\pm}(-1/\tau)}\exp\!\left(\!-\alpha c_3^\text{NS/R}+\frac{2\pi i}{\tau}c_1^\text{NS/R}\!\right)\!\prod_{k\geq0}\!\left(\!1{\pm}\exp\!\left(\!\alpha k^3-\frac{2\pi i}{\tau}k\!\right)\!\right)\!\!\!
\;,
\end{eqnarray}
is real when $\tau$ is pure imaginary. This shows that our results
are asymptotic rather than exact.

\subsection{Transformation under $\Gamma_1$}
\label{sec:3.3}
We will now look at the transformation of the expectation values under
the full modular group $\Gamma_1$=SL(2,$\mathbb{Z}$). Recall the
functions $F^{\text{NS/R,}\pm}_{2k}(\tau)$ defined in 
\eqref{eq:2.2.15}, these can be written as a combination of Eisenstein
series using the relations
(\ref{eq:2.2.12}--\ref{eq:2.2.14}). Under the modular transform 
$\hat\tau = \frac{-1}{\tau}$ we have 
\begin{eqnarray}
E_{2k}\left(\frac{-1}{\tau}\right) &=& \tau^{2k}E_{2k}(\tau)
\;,
\label{eq:3.3.1}
\\
E_{2k}\left(\frac{-2}{\tau}\right) &=&
2^{-2k}\tau^{2k}E_{2k}\left(\frac{1}{2}\tau\right)
\;,
\label{eq:3.3.2}
\\
E_{2k}\left(\frac{-1}{2\tau}\right) &=& 2^{2k}\tau^{2k}E_{2k}(2\tau)
\;,
\label{eq:3.3.3}
\end{eqnarray}
and hence we have
\begin{eqnarray}
F_{2k}^{\text{R,}+}\left(\frac{-1}{\tau}\right)&=&\tau^{2k}F_{2k}^{\text{NS,}-}(\tau)
\;,
\label{eq:3.3.4}
\\
F_{2k}^{\text{NS,}+}\left(\frac{-1}{\tau}\right)&=&\tau^{2k}F_{2k}^{\text{NS,}+}(\tau)
\;,
\label{eq:3.3.5}
\\
F_{2k}^{\text{NS,}-}\left(\frac{-1}{\tau}\right)&=&\tau^{2k}F_{2k}^{\text{R,}+}(\tau)
\;.
\label{eq:3.3.6}
\end{eqnarray}
The calculations of previous sections can be carried through to find that under the modular transform $\hat\tau=\frac{-1}{\tau}$
\begin{eqnarray}
\langle e^{\alpha I_{2m-1}}\rangle^{\text{R,}+}\left(\frac{-1}{\tau}\right)&\sim&\langle
e^{\sum_{n=1}\alpha^{(m)}_{2n(m-1)+1}
  I_{2n(m-1)+1}}\rangle^{\text{NS,}-}(\tau)
\;,
\label{eq:3.3.7}
\\
\langle e^{\alpha
  I_{2m-1}}\rangle^{\text{NS,}+}\left(\frac{-1}{\tau}\right)&\sim&\langle
e^{\sum_{n=1}\alpha^{(m)}_{2n(m-1)+1}
  I_{2n(m-1)+1}}\rangle^{\text{NS,}+}(\tau)
\;,
\label{eq:3.3.8}
\\
\langle e^{\alpha
  I_{2m-1}}\rangle^{\text{NS,}-}\left(\frac{-1}{\tau}\right)&\sim&\langle
e^{\sum_{n=1}\alpha^{(m)}_{2n(m-1)+1}
  I_{2n(m-1)+1}}\rangle^{\text{R,}+}(\tau)
\;,
\label{eq:3.3.9}
\end{eqnarray}
where the $\alpha^{(m)}_{2n(m-1)+1}$ are defined in \eqref{eq:3.2.4}.

Now look at the modular transform $\hat\tau=\tau+1$. Using the
expression \eqref{eq:3.1.2} one can see how the different sectors will
map into each other. We can also check that this agrees with our
expressions in terms of the Eisenstein series. Under this modular
transform only $E_{2k}(\frac{1}{2}\tau)$ transforms in a non trivial
way  
\begin{equation}
E_{2k}\left(\frac{1}{2}(\tau+1)\right)=(2^{2k}+2)E_{2k}(\tau)-E_{2k}\left(\frac{1}{2}\tau\right)-2^{2k}E_{2k}(2\tau)
\;.
\label{eq:3.3.10}
\end{equation}
Using the above transformation we find
\begin{eqnarray}
F_{2k}^{\text{R,}+}(\tau+1)&=&F_{2k}^{\text{R,}+}(\tau)
\;,
\label{eq:3.3.11}
\\
F_{2k}^{\text{NS,}+}(\tau+1)&=&F_{2k}^{\text{NS,}-}(\tau)
\;,
\label{eq:3.3.12}
\\
F_{2k}^{\text{NS,}-}(\tau+1)&=&F_{2k}^{\text{NS,}+}(\tau)
\;,
\label{eq:3.3.13}
\end{eqnarray}
and hence
\begin{eqnarray}
\langle e^{\alpha I_{2m-1}}\rangle^{\text{R,}+}(\tau+1)&=&\langle
e^{\alpha I_{2m-1}}\rangle^{\text{R,}+}(\tau)
\;,
\label{eq:3.3.14}
\\
\langle e^{\alpha I_{2m-1}}\rangle^{\text{NS,}+}(\tau+1)&=&\langle
e^{\alpha I_{2m-1}}\rangle^{\text{NS,}-}(\tau)
\;,
\label{eq:3.3.15}
\\
\langle e^{\alpha I_{2m-1}}\rangle^{\text{NS,}-}(\tau+1)&=&\langle
e^{\alpha I_{2m-1}}\rangle^{\text{NS,}+}(\tau)
\;,
\label{eq:3.3.16}
\end{eqnarray}
as expected. Note that for $\hat\tau=\tau+1$ we can see that we get an equality rather than expressions that just match as asymptotic series. We can see from the above expressions that the different sectors are transformed into each other in the same way as the characters of the free fermions without the charges. 

\subsection{Summary of results for $m=2$}

It will be helpful to collect here the results we have found for the
case of the single charge $I_3$ and the modular transform under the
generator $S: \tau \mapsto -1/\tau = \hat\tau$.

In each case, the first line gives the exact expression from the
bilinear form of the conserved charge, and the second line gives
an expression which is asymptotically equivalent for small $\alpha$.
\begin{align}
\Tr_{\NS+}\left(\hq^{L_0 - c/24} e^{\alpha I_3}\right)
&= \hq^{-1/48} e^{\frac{7}{1920}\alpha} \prod_{k=1/2}(1 + \hq^k e^{\alpha k^3})
\label{eq:NS+e}
\\
&\sim e^{\frac{\tau}{2 \pi i} \int_0^\infty \frac{t}{e^t+1}
  f(-\alpha\tau^3t^2/4) \rd t}
  \prod_{k=1/2} \left (1 + e^{2\pi i \tau k f(\alpha\tau^3k^2)}
  \right)\;,
\label{eq:NS+a}
\\
\Tr_{\R+}\left(\hq^{L_0 - c/24} e^{\alpha I_3}\right)
&= \hq^{1/24} e^{-\frac{1}{240}\alpha} \prod_{k=0}(1 + \hq^k e^{\alpha k^3})
\label{eq:R+e}
\\
&\sim 2^{1/2}\cdot e^{\frac{\tau}{2 \pi i} \int_0^\infty \frac{t}{e^t+1}
  f(-\alpha\tau^3t^2/4) \rd t}
  \prod_{k=1/2} \left (1 - e^{2\pi i \tau k f(\alpha\tau^3k^2)}
  \right)\;,
\label{eq:R+a}
\\
\Tr_{\NS-}\left(\hq^{L_0 - c/24} e^{\alpha I_3}\right)
&=  \hq^{-1/48} e^{\frac{7}{1920}\alpha} \prod_{k=1/2}(1 - \hq^k e^{\alpha k^3})
\label{eq:NS-e}
\\
&\sim 2^{-1/2}\cdot e^{-\frac{\tau}{2 \pi i} \int_0^\infty \frac{t}{e^t-1}
  f(-\alpha\tau^3t^2/4) \rd t}
  \prod_{k=0} \left (1 + e^{2\pi i \tau k f(\alpha\tau^3k^2)}
  \right)\;,
\label{eq:NS-a}
\end{align}
where $f(z)=f_2(z)$ defined in \eqref{eq:3.2.15}.

The asymptotic expressions are those one would get as a trace over the
appropriate space
\be
\Tr_{\NS/R,\pm}( e^{2\pi i \tau\,H} )
=
\Tr_{\NS/R,\pm}( q^{L_0 - c/24} e^Q )
\;,
\ee
with
\be
   H = h_0^{\NS/\R} + \sum_k k f(\alpha \tau k^2) \psi_{-k}\psi_k
\;,\;\;\;\;
\ee\be
 h_0^{\NS} = -\frac{1}{4 \pi^2} \int_0^\infty \frac{t}{e^t+1}
  f(-\alpha\tau^3t^2/4\pi^2) \rd t
\;,\;\;\;\;
  h_0^{\R} = \frac{1}{4 \pi^2}\int_0^\infty \frac{t}{e^t-1}
  f(-\alpha\tau^3t^2/4\pi^2) \rd t
\;.
\label{eq:h0def}
\ee

We will propose exact formulae later, but first we will check our
asymptotic results against previously-known results.

\section{Comparison with known results}
\label{sec:Dijk}
There are two checks we can now do on our results. The first is to see
if our coefficients $\alpha^{(n)}_{2n(m-1)+1}$ match those we would
find using the results from \cite{Dijkgraaf}. \cite{Dijkgraaf} uses
the pre-Lie algebra structure mentioned in section \ref{sec:2} and
tells us how to map between a GGE where the insertions are either
integrals over the whole torus or just the integral over one
cycle. This allows us to map to the case where the integrals are over
the whole torus which makes it easier to perform the modular transform
and then map back to the charges we have been using. 

We will also use
the results from \cite{Ross} which give the expectation values of
products of the charges in terms of a modular differential operator
acting on the character. These results can be compared with the
coefficients in our asymptotic expansion which give the modular
transform of the expectation values $\langle
I_{2m-1}^n\rangle^{\text{NS/R},\pm}(\hat\tau)$. 

In both cases our
results match the previously found results.

\subsection{Dijkgraaf Master Equation}
\label{sec:DME}

We will now make our first comparison with results from the literature. We will use the methods from \cite{Dijkgraaf} to find an alternate way of calculating the $\alpha^{(m)}_{2n(m-1)+1}$ of \eqref{eq:3.2.4} and see that the results match. To do this we will use the Master equation from \cite{Dijkgraaf}. To start, we want to find a relationship between the constants $t_n$ and $s_n$ such that
\begin{equation}
\langle\exp\left(\sum_{n=1}\left(\int t_nJ_{2n}\right)\right)\rangle^{\text{NS/R},\pm} =
\langle\exp\left(\sum_{n=1}\left(\oint s_nJ_{2n}\right)\right)\rangle^{\text{NS/R},\pm}
\;,
\label{eq:4.1.1}
\end{equation}
where 
\begin{eqnarray}
\int t_nJ_{2n}&=&\int\frac{d^2z}{2\pi\tau_2}J_{2n}(z)
\;,
\label{eq:4.1.2}
\\
\oint t_nJ_{2n}&=&\int_0^1\frac{\rd z}{2\pi}J_{2n}(z)
\;.
\label{eq:4.1.3}
\end{eqnarray}
These integrals are different from our previous definitions, it will be easier to use these expressions in this section as they are used by Dijkgraaf and we will explain how to relate them to our previous integrals at the end of this section.  The relation between the $t_n$ and $s_n$ will depend on $\tau$ so we can use it to find the modular transform of $\langle e^{\alpha I_{2m-1}}\rangle^{\text{NS/R},\pm}$. We will again use that the expectation value of the quasi-primary field of weight $2n$ is a weight $2n$ modular form and it's generalisation to higher point functions
\begin{equation}
\left\langle\int J_{2n_1}\dots\int J_{2n_I}\right\rangle^{\text{NS/R},\pm}(\hat\tau)=(c\tau+d)^{2\sum_{i=1}^I n_i}\left\langle\int J_{2n_1}\dots\int J_{2n_I}\right\rangle^{\text{NS/R},\pm}(\tau)
\;.
\label{eq:4.1.4}
\end{equation}
Since we are integrating over the torus the ordering of the fields $J_{2n}$ doesn't matter, we have the modular transform relation
\begin{eqnarray}
&\left\langle\exp\left(\sum_{n=1}\left(\oint s_nJ_{2n}\right)\right)\right\rangle^{\text{NS/R},\pm}(\hat\tau)\nonumber\\
&=\left\langle\exp\left(\sum_{n=1}\left(\int t_n^{(\hat\tau)}(s_m)J_{2n}\right)\right)\right\rangle^{\text{NS/R},\pm}(\hat\tau)\nonumber\\
&=\left\langle\exp\left(\sum_{n=1}\left(\int
 (c\tau{+}d)^{2n}t_n^{(\hat\tau)}(s_m)J_{2n}\right)\right)\right\rangle^{\text{NS/R},\pm}(\tau)\nonumber\\
&=\left\langle\exp\left(\sum_{n=1}\left(\oint
  s_n^{(\tau)}((c\tau{+}d)^{2m}t_m^{(\hat\tau)}(s_l))J_{2n}\right)\right)\right\rangle^{\text{NS/R},\pm}(\tau)
\;.
\label{eq:4.1.5}
\end{eqnarray}
where the superscripts $\tau$ and $\hat\tau$ indicate whether we use $\tau$ or $\hat\tau$ in the relations between $s_n$ and $t_n$. If such a relation can be found so \eqref{eq:4.1.5} holds this can be used to justify our asymptotic expansion of $Q_m$ in \eqref{eq:3.2.2}.

In order to find the relation between the $t_n$ and $s_n$ we will copy the method from \cite{Dijkgraaf}. Define
\begin{equation}
Z[s,t]=\langle\exp\left(\sum_{n=1}\left(\int t_nJ_{2n}+\oint
s_nJ_{2n}\right)\right)\rangle^{\text{NS/R},\pm}
\;,
\label{eq:4.1.6}
\end{equation}
where $s= \{s_1,\dots\}$ and $t=\{t_1,\dots\}$. We want to find the relation between $s_n$ and $t_n$ that gives $Z[s,0]=Z[0,t]$. Now define
\begin{equation}
a_n=\frac{s_n}{2\tau_2}+\delta^{n,1} \text{ and }b_n=-\frac{t_n}{2\tau_2}+\delta^{n,1}
\;.
\label{eq:4.1.7}
\end{equation}
In \cite{Dijkgraaf} the relation \eqref{eq:4.1.7} contains $\delta^{n,T}$ instead, where $T$ labels the stress tensor. Here $T=J_2$ so $\delta^{n,T}=\delta^{n,1}$. Now if we consider $Z$ as a function of $a_n, b_n$ instead of $s_n, t_n$ it is shown in \cite{Dijkgraaf} that $Z[a,b]$ obeys the differential equation
\begin{equation}
(L_n^{(a)}+L_n^{(b)})Z[a,b]=0\text{ for }n=1,2,\dots
\;,
\label{eq:4.1.8}
\end{equation}
where
\begin{equation}
L_n^{(a)}=\sum_{m=1}(2m-1)a_m\frac{\partial}{\partial a_{n+m-1}}\text{ for }n=1,2,\dots
\;,
\label{eq:4.1.9}
\end{equation}
and similarly for $L_n^{(b)}$.

As mentioned in section \ref{sec:2}, the fields $J_{2n}$ generate a pre-Lie algebra, W, as defined in \cite{Dijkgraaf}. We have $Z:W\times W\rightarrow \mathbb{C}$ and, from section \ref{sec:2}, $W$ is isomorphic to the space of odd holomorphic functions. Define the odd holomorphic functions
\begin{equation}
a(z)=\sum_{k=1}a_kz^{2k-1}\text{ and }b(z)=\sum_{k=1}b_kz^{2k-1}
\;.
\label{eq:4.1.10}
\end{equation}
Now $Z$ can be thought of as a complex function of $a(z)$ and $b(z)$. $L_n^{(a)}$ acts on $a(z)$ by
\begin{equation}
L_n^{(a)}a(z)=\sum_{m=1}(2m-1)a_mz^{2(m+n-1)-1}
\;.
\label{eq:4.1.11}
\end{equation}
This is equivalent to the action of
\begin{equation}
L_n=z^{2n-1}\frac{\partial}{\partial z}
\;,
\label{eq:4.1.12}
\end{equation}
so we have
\begin{equation}
L_na(z)=L_n^{(a)}a(z)
\;.
\label{eq:4.1.13}
\end{equation}
The $L_n$'s generate odd holomorphic functions. The $L_n$ and $L_n^{(b)}$ both act on $b(z)$ in the same way. From \eqref{eq:4.1.8} we can see that $Z[a,b]$ is invariant under diffeomorphisms generated by the $L_n$, these are odd holomorphic functions. This means 
\begin{equation}
Z[a(f(z)),b(f(z))]=Z[a(z),b(z)]
\;,
\label{eq:4.1.14}
\end{equation}
where $f(z)$ is any odd holomorphic function. Set $f(z)=b^{-1}(z)$ ($b(z)$ is odd so $b^{-1}(z)$ is odd) and $a(z)=z$ to get
\begin{equation}
Z[z,b(z)]=Z[b^{-1}(z),z]
\;.
\label{eq:4.1.15}
\end{equation}
The relations between $a_n, s_n$ and $b_n, t_n$ in \eqref{eq:4.1.7} then allows us to write \eqref{eq:4.1.15} as $Z[0,t]=Z[s,0]$ as required. From $a(b(z))=z$ we have the following relation between $t$ and $s$ 
\begin{eqnarray}
w=\sum_{n=1}a_nz^{2n-1},&& z=\sum_{n=1}b_nw^{2n-1}
\;,
\label{eq:4.1.16}
\\
\Rightarrow w=z+\sum_{n=1}\frac{s_n}{2\tau_2}z^{2n-1},&&z=w-\sum_{n=1}\frac{t_n}{2\tau_2}w^{2n-1}
\;.
\label{eq:4.1.17}
\end{eqnarray}
We can now use these relations to find $s_n$ in terms of $t_n$, we set $s_1=0=t_1$ (they are related by $s_1=\frac{2\tau_2t_1}{2\tau_2-t}$). The first few of these relations are
\begin{eqnarray}
s_2&=&t_2
\;,
\label{eq:4.1.18}
\\
s_3&=&t_3+\frac{3t_2^2}{2\tau_2}=t_3+\frac{3s_2^2}{2\tau_2}
\;,
\label{eq:4.1.19}
\\
s_4&=&t_4+\frac{4t_2t_3}{\tau_2}+\frac{3t_2^3}{\tau_2^2}=t_4+\frac{4s_2s_3}{\tau_2}-\frac{3s_2^3}{\tau_2^2}
\;.
\label{eq:4.1.20}
\end{eqnarray}
These can be used to find the coefficients $s_n^{(\tau)}$ from \eqref{eq:4.1.5}. Considering the case where $s_n=0$ for $n\neq2$ the relations (\ref{eq:4.1.18}--\ref{eq:4.1.20}) can be used to find
\begin{eqnarray}
t_2^{(\hat\tau)}&=&s_2
\;,
\label{eq:4.1.21}
\\
t_3^{(\hat\tau)}&=&-\frac{3s_2^2}{2\hat\tau_2}
\;,
\label{eq:4.1.22}
\\
t_4^{(\hat\tau)}&=&\frac{3s_2^3}{\hat\tau_2^2}
\;,
\label{eq:4.1.23}
\\
s_2^{(\tau)}&=&(c\tau+d)^4s_2
\;,
\label{eq:4.1.24}
\\
s_3^{(\tau)}&=&3ic(c\tau+d)^7s_2^2
\;,
\label{eq:4.1.25}
\\
s_4^{(\tau)}&=&-12c^2(c\tau+d)^{10}s_2^3
\;.
\label{eq:4.1.26}
\end{eqnarray}
We need to relate the $s_i$ and $\alpha^{(m)}_{2n+1}$ coefficients. Recall from \eqref{eq:2.2.1}
\begin{equation}
I_{2n-1}=\int_0^{2\pi i}\frac{\rd w}{2\pi i}J_{2n}(w)
\;.
\label{eq:4.1.27}
\end{equation}
Let $w=2\pi i\tilde w$, $J_{2n}$ is a quasi-primary field of weight $2n$ so $J_{2n}(\tilde w) = (2\pi i)^{2n}J_{2n}(w)$. We then have
\begin{equation}
I_{2n-1}=\int_0^1\frac{d\tilde w}{(2\pi i)^{2n}}J_{2n}(\tilde
w)=-i(2\pi i)^{1-2n}\int J_{2n}
\;.
\label{eq:4.1.28}
\end{equation}
Hence we should identify $\alpha=i(2\pi i)^{2m-1}s_m$ and
$\alpha^{(m)}_{2n+1}=i(2\pi i)^{2n+1}s_{n+1}^{(\tau)}$ when comparing
the results of section \ref{sec:3.2} 
to the results from this section. For $n=1,\dots,13$ we have found
$s_n^{(\tau)}$ for the cases $m=2,\dots,13$ and verified that using
the above relations between $\alpha, s_m, \alpha_{2n+1}^{(m)}$ and
$s_{n+1}^{(\tau)}$ we get the same results as we did in section
\ref{sec:3.2}.

\subsection{$\langle I_{2m-1}^n\rangle^{\text{NS/R},\pm}$ as a modular differential
  operator acting on a character}
\label{sec:4.2}

We can also compare our results with those in \cite{Ross}. \cite{Ross}
gives expressions for the expectation values of the charges in the
form of differential operators acting on the characters of the
representations. The normalisation of the charges in \cite{Ross} is
different from the conventions used here and the expectation values of
the charges are defined without the factor of
$\frac{1}{\chi^{\text{NS/R},\pm}(\tau)}$. If we denote the charges from \cite{Ross} by
$\tilde I_{2n-1}$ then the relation between $\langle I_{n_1}\dots
I_{n_i}\rangle^{\text{NS/R},\pm}$ and $\langle \tilde I_{n_1}\dots\tilde I_{n_i}\rangle^{\text{NS/R},\pm}$ is  
\begin{equation}
\langle I_{n_1}\dots I_{n_i}\rangle^{\text{NS/R},\pm}=\frac{A_{n_1}\dots A_{n_i}}{\chi^{\text{NS/R},\pm}(\tau)}\langle\tilde I_{n_1}\dots\tilde I_{n_i}\rangle^{\text{NS/R},\pm}
\;,
\label{eq:4.2.1}
\end{equation}
where $A_{n_i}$ are numerical constants that can be determined by finding the difference in normalisation used when defining the quasi-primary fields
\begin{equation}
\langle I_{2m-1}\rangle^{\text{NS/R},\pm}=\frac{A_{2m-1}}{\chi^{\text{NS/R},\pm}(\tau)}\langle\tilde I_{2m-1}\rangle^{\text{NS/R},\pm}
\;.
\label{eq:4.2.2}
\end{equation}
The differential operators that act on the characters to give the
expectation values of the $\tilde I_{2n-1}$ can be found in
\cite{Ross}, for the free fermions we must set
$k=\frac{c}{24}=\frac{1}{48}$. The expectation values are quasi-modular forms so for each case we want to check, if the
expressions match up to a certain order in $q=e^{2\pi i\tau}$ (the
order depends on the weight and depth) then they are equal. (See appendix \ref{app:mf} for the definition of the space of quasi-modular forms of weight $k$ and depth $p$, $\tilde M_k^{(\leq p)}(\Gamma(2))$). We will
first compare the one point functions with the results from
\cite{Ross} to determine the $A_{2m-1}$. For $\langle I_3\rangle^{\text{NS/R},\pm},
\langle I_5\rangle^{\text{NS/R},\pm}, \langle I_7\rangle^{\text{NS/R},\pm}$ and $\langle I_9\rangle^{\text{NS/R},\pm}$ we
find that the results match if $A_3=\frac{6}{7}, A_5=\frac{144}{143},
A_7=\frac{323}{432}$ and $A_9=\frac{10925}{20736}$. If we then look at
$\langle I_3^2\rangle^{\text{NS/R},\pm}$ and $\langle I_5^2\rangle^{\text{NS/R},\pm}$ we find the results
again match if the numerical constants are $A_3^2$ and $A_5^2$
respectively, as expected.

We can easily see from our results and those in \cite{Ross} that the
one point functions transform as modular forms of the correct
weight. We will now compare the modular transform of $\langle
I_3^2\rangle^{\text{NS/R},\pm}, \langle I_5^2\rangle^{\text{NS/R},\pm}$ and $\langle I_3^3\rangle^{\text{NS/R},\pm}$, which
come from the asymptotic expansion of \eqref{eq:3.2.5}, with $\langle\tilde
I_3^2\rangle^{\text{NS/R},\pm}, \langle\tilde I_5^2\rangle^{\text{NS/R},\pm}$ and $\langle\tilde
I_3^3\rangle^{\text{NS/R},\pm}$ which follow from the modular transform of the
differential operators acting on the characters. For $\langle
I_3^2\rangle^{\text{NS/R},\pm}$ the results match if the following modular form vanishes 
\begin{equation}
(55296D^3-2568E_4(\tau)D+23E_6(\tau))\chi^{\text{NS/R},\pm}(\tau)=0
\;,
\label{eq:4.2.3}
\end{equation}
where $D^n=D_{2n-2}D_{2n-4}\dots D_2D_0$, $D_r=\frac{1}{2\pi
  i}\frac{d}{d\tau}-\frac{r}{12}E_2(\tau)$ is the Serre derivative and $E_{2n}(\tau)$ is the
Eisenstein series of weight $2n$.
The left hand side of \eqref{eq:4.2.3} is a
modular form in $M_6(\Gamma(2))$. This space of modular forms is 4
dimensional and the left hand side of \eqref{eq:4.2.3} vanishes up to order
$q^3$ hence it vanishes identically. So we have agreement between the
results derived here and the results from \cite{Ross} . For $\langle I_5^2\rangle^{\text{NS/R},\pm}$ we find a weight 10 modular form that must vanish
\begin{equation}
(1560674304D^5{+}40048128E_4D^3{+}48968640E_6D^2{-}17954904E_4^2D{+}263189E_4E_6)\chi^{\text{NS/R},\pm}(\tau){=}0
\;.
\label{eq:4.2.4}
\end{equation}
The space $M_{10}(\Gamma(2))$ is 6 dimensional and (8.4) vanishes up to order $q^5$ hence \eqref{eq:4.2.4} holds. Finally for $\langle I_3^3\rangle^{\text{NS/R},\pm}$ we find that there is a modular form of weight 8 and a quasi-modular form of weight 10 and depth 1 that must vanish
\begin{eqnarray}
&(110592D^4-5136E_4D^2+1758E_6D-23E_4^2)\chi^{\text{NS/R},\pm}(q)=0
\;,
\label{eq:4.2.5}
\\
&(159252480D^5+19201536E_4D^3+4996800E_6D^2-2534088E_4^2D+33143E_4E_6+\nonumber\\
&E_2(159252480D^4-7395840E_4D^2+2531520 E_6D-33120E_4^2))\chi^{\text{NS/R},\pm}(q)=0
\;.
\label{eq:4.2.6}
\end{eqnarray}
The left hand side of \eqref{eq:4.2.5} vanishes up to order $q^4$ so \eqref{eq:4.2.5} holds. The left hand side of \eqref{eq:4.2.6} lives in the space of quasi-modular forms of weight 10 and depth 1, $\tilde M_k^{(\leq 1)}(\Gamma(2))$, which is 11 dimensional. The left hand side of \eqref{eq:4.2.6} vanishes up to order 10 so \eqref{eq:4.2.6} holds. Hence our asymptotic results match the results from \cite{Ross} in the above cases.

\section{Thermodynamic Bethe Ansatz}
\label{sec:TBA}

An alternative method to find the partition function in the crossed
channel is the Thermodynamic Bethe Ansatz (TBA). 
The original use of the TBA \cite{ZTBA}
was to find the ground state energy of a
finite-size system by considering the expansion of the partition
function of a finite-size system in the two channels.
In the first channel, it is given as a sum over a large number of
particles with a known scattering matrix. By minimisation of the
free energy, this leads to a closed form for the finite-size ground-state
energy in the opposite channel. This was further extended to a
construction of the full set of excited-state energies by analytic
continuation of the ground state energy \cite{Dorey}.

We can use this method in our case by modifying the  ``Energy''
of the particles in the scattering channel by adding a coupling to the
KdV charges through the chemical potential.
The resulting ``ground state energy'' expression in the
second channel should give the ground-state value of the full
Hamiltonian in the crossed channel.

We will limit ourselves (again) to a coupling to a single charge
$I_{2m-1}$ (with chemical potential $\alpha$)
We will find in section \ref{sec:5.1}  that naively modifying the Thermodynamic Bethe
Ansatz gives us the same ground state energy as the conjectured
integrals 
\eqref{eq:Rgnd} and \eqref{eq:NSgnd} in the case $(c\tau+d) =-\tau$.

In section \ref{sec:5.2b} we look at the singularities in the TBA
system that can correspond to excited state energies in the simplest
case of $m=2$, that is the simplest charge $I_3$ and recover the 
known contributions 
\eqref{eq:3.2.17}
and two further sets.

Throughout this section we will work with a
rectangular
torus so the modular parameter is $\tau=i\tau_2$. The modular
parameter is the ratio of the two periods of the torus, here the two
cycles have length $L$ and $R$ where $L$ is acting as the time
dimension and $R$ as the spatial dimension so $\tau_2=\frac{L}{R}$.

\subsection{Modified thermodynamic Bethe ansatz}
\label{sec:5.1}
The derivation of the TBA equations for a system where the total energy and momentum are sums over the one particle energy and momentum respectively can be found in \cite{ZTBA} and in \cite{KMTBA}, and in \cite{Fendley:1991xn} the TBA with a chemical potential is derived. We will use the same derivation here but instead of the total energy and momentum being sums over the one particle energies and momenta we take them to be the sum over a function of the one particle energy and momentum. The on shell energy and momenta for a particle will be denoted by $e(\theta)$ and $p(\theta)$ respectively where $\theta$ is the rapidity of the particle. If we have massive particles of mass $M$ then as in \cite{ZTBA} and \cite{KMTBA} the on shell energy and momentum are
\begin{eqnarray}
e(\theta)&=&M\cosh(\theta)
\;,
\label{eq:5.1.1}
\\
p(\theta)&=&M\sinh(\theta)
\;.
\label{eq:5.1.2}
\end{eqnarray}
If instead we have massless particles then, as in \cite{Fendley}, we have a mass scale $M$ and the energy and momentum are
\begin{equation}
e(\theta)=\pm p(\theta)=\frac{1}{2}Me^{\pm\theta}
\;,
\label{eq:5.1.3}
\end{equation}
where the $+$ is for right movers and the $-$ is for left movers. Consider a system where the energy and momentum of a particle is a function of their on shell energy and momentum only, there are no contributions from the other particles. The one particle energy is $\mathcal{E}(e(\theta))$ and the one particle momentum is $\mathcal{P}(p(\theta))$. If there are $N$ particles with rapidities $\theta_i$, $i=1,\dots,N$, then the total energy and momentum are
\begin{eqnarray}
H&=&\sum_{i=1}^N\mathcal{E}(e(\theta_i))
\;,
\label{eq:5.1.4}
\\
P&=&\sum_{i=1}^N\mathcal{P}(p(\theta_i))
\;.
\label{eq:5.1.5}
\end{eqnarray}
Following through the method in \cite{Fendley:1991xn} but with our total energy and momentum we find the integral equation for $\epsilon(\theta)$
\begin{equation}
R\mathcal{E}(e(\theta))-\epsilon(\theta)-\int_{-\infty}^\infty\frac{d\theta'}{2\pi}\varphi(\theta-\theta')\log(1+ \lambda e^{-\epsilon(\theta')})=0
\;,
\label{eq:5.1.6}
\end{equation}
where $\lambda$ is the fugacity which is related to the chemical potential $\mu$ via $\lambda=e^{R\mu}$ and $\varphi(\theta)=-i\frac{d}{d\theta}S(\theta)$ with $S(\theta)$ the S matrix of the system. The free energy $f(R)$ is given by
\begin{equation}
Rf(R)=-\int_{-\infty}^\infty \frac{d}{d\theta}\mathcal{P}(p(\theta))\log(1+\lambda e^{-\epsilon(\theta)})\frac{d\theta}{2\pi}
\;.
\label{eq:5.1.7}
\end{equation}
For a free theory we have $\varphi=0$ so
\begin{equation}
\epsilon(\theta)=R\mathcal{E}(e(\theta))
\;,
\label{eq:5.1.8}
\end{equation}
and \eqref{eq:5.1.7} becomes
\begin{equation}
Rf(R)=-\int_{-\infty}^\infty \frac{d}{d\theta}\mathcal{P}(p(\theta))\log(1+\lambda e^{-R\mathcal{E}(e(\theta))})\frac{d\theta}{2\pi}
\;.
\label{eq:5.1.9}
\end{equation}
First let us consider the case of massive particles. We take
$e(\theta)$ and $p(\theta)$ as in \eqref{eq:5.1.1} and
\eqref{eq:5.1.2} and then using the substitution $x=MR\cosh(\theta)$
in \eqref{eq:5.1.9} we find 
\begin{eqnarray}
Rf(R)&=&-2\int_{MR}^\infty\frac{dx}{2\pi}\left(\frac{d}{dx}\mathcal{P}\left(\sqrt{\frac{x^2-(MR)^2}{R^2}}\right)\right)\log\left(1+\lambda
e^{-R\mathcal{E}\left(\frac{x}{R}\right)}\right)\;,
\label{eq:5.1.10}
\\
&=&-\frac{2}{R}\int_0^\infty\frac{dx}{2\pi}\mathcal{P}'\left(\frac{x}{R}\right)\log\left(1+\lambda e^{-R\mathcal{E}\left(\frac{x}{R}\right)}\right)\text{ as }M\rightarrow0
\;.
\label{eq:5.1.11}
\end{eqnarray}
Alternately if we have massless right moving particles with
$e(\theta)$ and $p(\theta)$ as in \eqref{eq:5.1.3} and use the
  substitution $x=\frac{1}{2}MRe^\theta$ in \eqref{eq:5.1.9} we get 
\begin{equation}
Rf(R)=-\frac{1}{R}\int_0^\infty\frac{dx}{2\pi}\mathcal{P}'\left(\frac{x}{R}\right)\log\left(1+\lambda e^{-R\mathcal{E}\left(\frac{x}{R}\right)}\right)
\;.
\label{eq:5.1.12}
\end{equation}
Note that the mass scale $M$ is no longer present in the integral. We
can see that \eqref{eq:5.1.11} and \eqref{eq:5.1.12} differ by a
factor of 2 which is expected since the massless limit of \eqref{eq:5.1.10}
contains both left and right movers.

We can integrate \eqref{eq:5.1.12} by parts to get
\be
Rf(R)=-\int_0^\infty\frac{dx}{2\pi}\frac{\lambda\mathcal{P}\left(\frac{x}{R}\right)\mathcal{E}'\left(\frac{x}{R}\right)}{e^{R\mathcal{E}\left(\frac{x}{R}\right)}+\lambda}\;.\label{eq:feip}
\ee
We have removed the branch point in the integrand but have instead introduced poles. We will use the position of these poles in section \ref{sec:5.2b} when we consider the possible contributions to the one particle energies. 

\subsection{Free energy}
\label{sec:5.2}

Rather than consider the general GGE with chemical potentials for all
the charges, we shall restrict to the case where there is
just one charge $I_{2m-1}$ with chemical potential $\alpha$.
In the thermodynamic limit $L\rightarrow\infty$ and hence so does
$\tau_2=\frac{L}{R}$. In our results we can see that $\alpha$ always
appears with $\tau_2$ in the combination $\alpha\tau_2^{2m-1}$. Hence
we will consider the limit where $\tau_2\rightarrow\infty$,
$\alpha\rightarrow0$ and $\alpha\tau_2^{2m-1}$ remains constant. We
will continue to write $\alpha\tau_2^{2m-1}$ to make comparison with
our earlier results easier. 

Including the charge $I_{2m-1}$ with chemical potential $\alpha$ 
corresponds to values for the momentum and energy
in the massless equation, \eqref{eq:5.1.12} 
\be
\mathcal{P}(x)=x\;,\;\;\; \mathcal{E}(x)=x-\frac{\alpha}{R}\left(\frac{R\tau_2}{2\pi}\right)^{2m-1}x^{2m-1}\;.
\ee
Substituting these values,
we find that the integrals have an asymptotic expansion
about $\alpha\tau_2^{2m-1}=0$ that is equal to the series \eqref{eq:3.2.8} 
\begin{eqnarray}
-\tau_2\int_0^\infty\frac{dx}{2\pi}\log\left(1+e^{-x+\alpha\left(\frac{\tau_2x}{2\pi}\right)^{2m-1}}\right)+2\pi\tau_2
c_1^{\text{NS}}&\sim&\sum_{n=1}\alpha^{(m)}_{2n(m-1)+1}c_{2n(m-1)+1}^{\text{NS}}\;,
\label{eq:5.2.1}
\\
-\tau_2\int_0^\infty\frac{dx}{2\pi}\log\left(1-e^{-x+\alpha\left(\frac{\tau_2x}{2\pi}\right)^{2m-1}}\right)+2\pi\tau_2
c_1^{\text{R}}&\sim&\sum_{n=1}\alpha^{(m)}_{2n(m-1)+1}c_{2n(m-1)+1}^{\text{R}}
\;,
\label{eq:5.2.2}
\end{eqnarray}
where $\lambda=1$ in the NS sector and $\lambda=-1$ in the R sector.

Using a change of variables we can explicitly show that the integral
\eqref{eq:5.2.1} matches the integral in \eqref{eq:NSgnd} and
\eqref{eq:5.2.2} matches the integral in \eqref{eq:Rgnd}. First consider the change of
variables $x=t\tilde f_m(t)$ where
\begin{equation}
\alpha\left(\frac{\tau_2}{2\pi}\right)^{2m-1}(t\tilde f_m(t))^{2m-1}-t\tilde f_m(t)+t=0
\;.
\label{eq:5.2.3}
\end{equation}
We will assume that $\tilde f_m(t)$ is real for $t>0$,
$\lim_{t\rightarrow0}t\tilde f_m(t)=0$,
$\lim_{t\rightarrow\infty}t\tilde f_m(t)=\infty$ and
$\lim_{t\rightarrow\infty}t\tilde f_m(t)\log(1+e^{-t})=0$. Using this change
of variables in \eqref{eq:5.2.1} and \eqref{eq:5.2.2} gives us
\begin{equation}
\int_0^\infty\frac{dx}{2\pi}\log\left(1\pm
e^{-x+\alpha\left(\frac{\tau_2x}{2\pi}\right)^{2m-1}}\right)=\pm\int_0^\infty\frac{dt}{2\pi}\frac{t}{e^t\pm1}\tilde f_m(t)
\;.
\label{eq:5.2.4}
\end{equation}
If we take 
\begin{equation}
\tilde f_m(t)=f_m\left(\frac{i\alpha \tau_2^{2m-1}t^{2m-2}}{(2\pi)^{2m-2}}\right)
\;,
\label{eq:5.2.5}
\end{equation}
where, $f_m$ is defined in \eqref{eq:3.2.15}, then from \cite{Glasser}
they satisfy \eqref{eq:5.2.3} and we have the same integrals as in
\eqref{eq:Rgnd} and \eqref{eq:NSgnd} in the case $c=-1,d=0$. 
This choice for the $\tilde f_m(t)$ will also satisfy the
above conditions in the $t\rightarrow0$ and $t\rightarrow\infty$
limits and for $\alpha<0$, $\tilde f_m(t)$ is real for all $t$.

\subsection{One particle energies}
\label{sec:5.2b}

In \cite{Dorey} it is shown how varying the parameter $r=MR$ into the
complex plane and encircling the singularities of the integrand allows
us to pick up the one particle energies. 

\begin{figure}
\centering
\begin{subfigure}[b]{0.4\textwidth}
  \begin{centering}
  \includegraphics[width=0.9\textwidth]{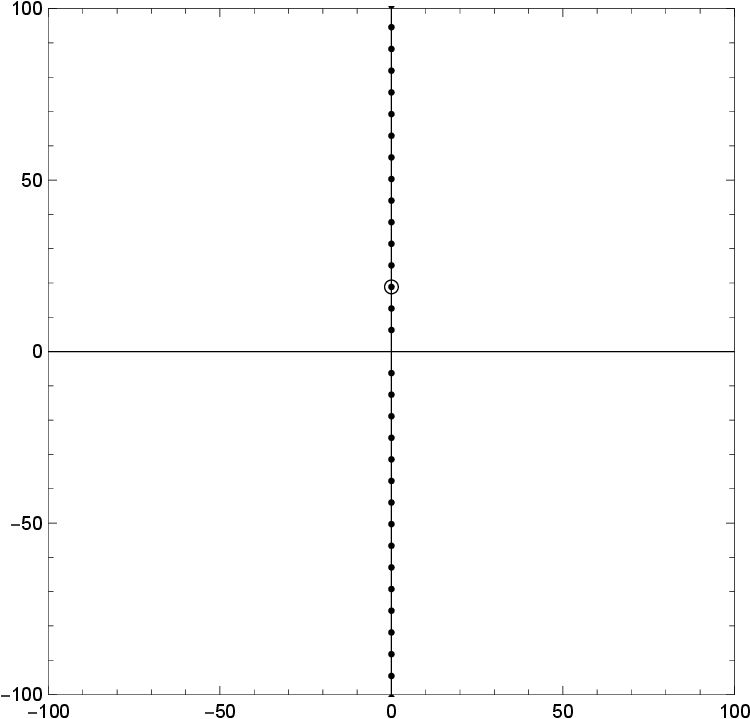}
  \end{centering}
\caption{$\gamma=-0.00001$}
\end{subfigure}
\begin{subfigure}[b]{0.4\textwidth}
  \begin{centering}
  \includegraphics[width=0.9\textwidth]{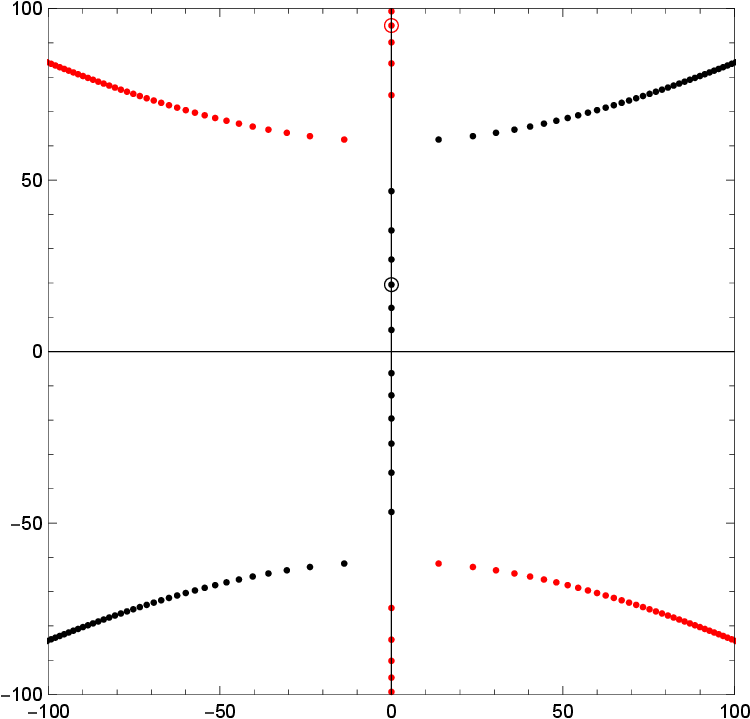}
  \end{centering}
\caption{$\gamma=-0.0204785$}
\end{subfigure}
\\
\begin{subfigure}[b]{0.4\textwidth}
  \begin{centering}
  \includegraphics[width=0.9\textwidth]{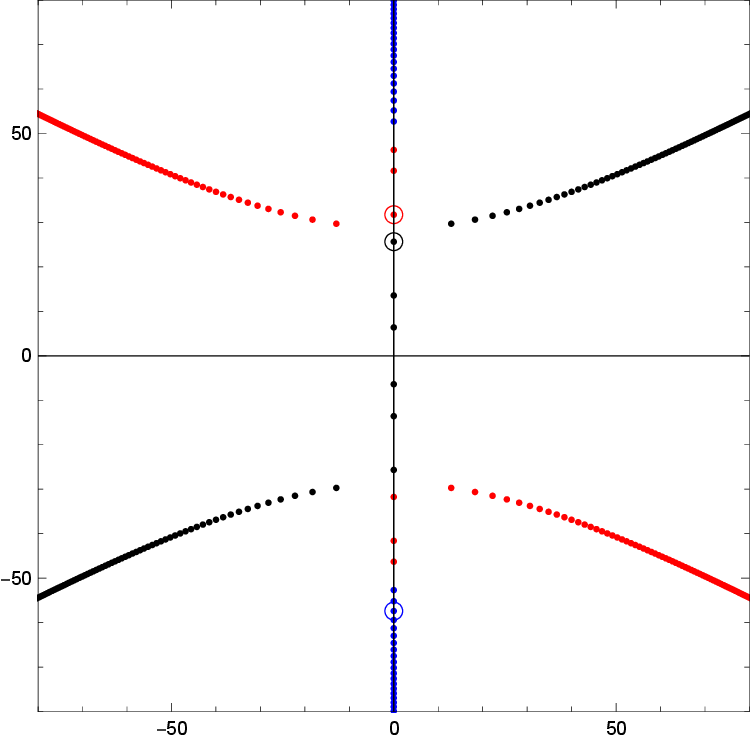}
  \end{centering}
\caption{$\gamma=-0.0930842$}
\end{subfigure}
\begin{subfigure}[b]{0.4\textwidth}
  \begin{centering}
  \includegraphics[width=0.9\textwidth]{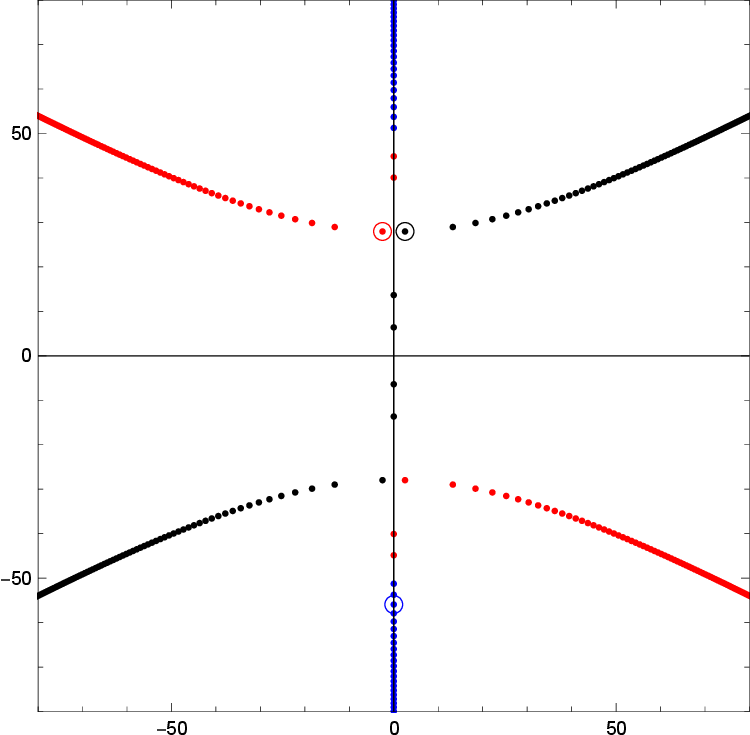}
  \end{centering}
\caption{$\gamma=-0.0986693$}
\end{subfigure}
\\
\begin{subfigure}[b]{0.4\textwidth}
  \begin{centering}
  \includegraphics[width=0.9\textwidth]{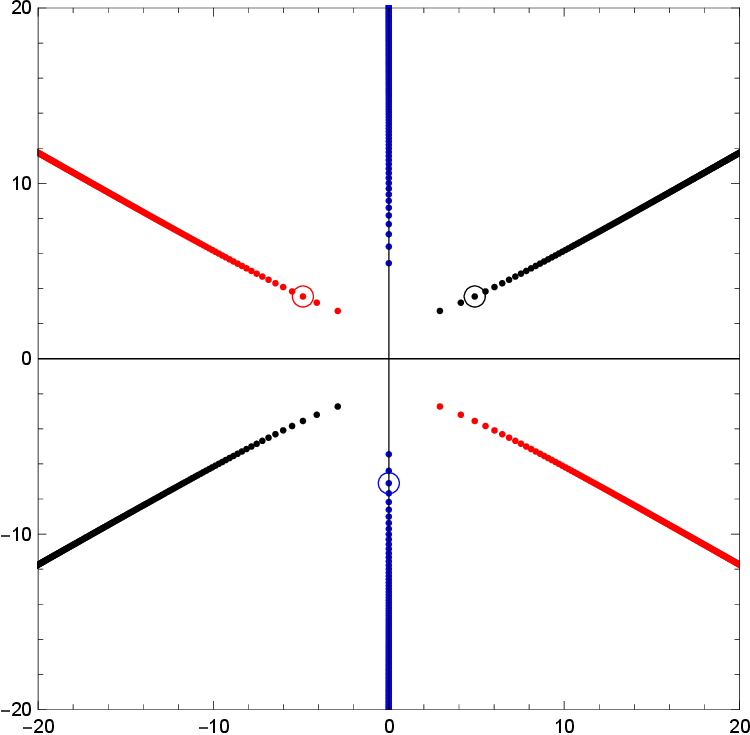}
  \end{centering}
\caption{$\gamma=-0.167552$}
\end{subfigure}
\begin{subfigure}[b]{0.4\textwidth}
  \begin{centering}
  \includegraphics[width=0.88\textwidth]{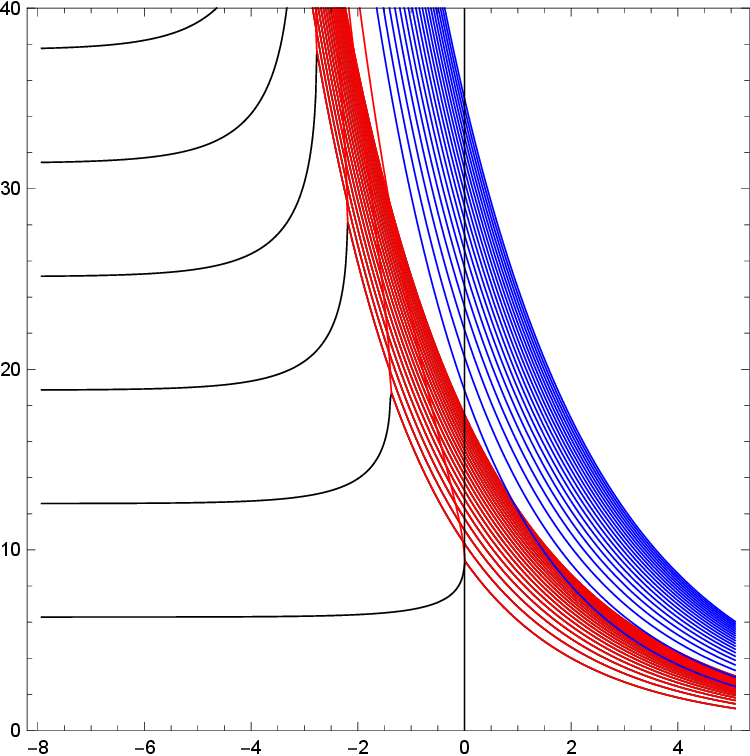}
  \end{centering}
\caption{$\ $}
\end{subfigure}
\caption{Figures (a)--(e) are plots of the roots $x_i(n)$ for various real values of $\gamma$
  and $n\in\mathbb Z$.
The black points are $x_1(n)$, the red are $x_2(n)$ and the blue,
$x_3(n)$. In each plot, the circled points correspond to
$n=3$. Note that the plots are not all scaled the same. Figure (f) is
a plot of the imaginary parts of the eigenvalues in the upper half
plane, plotted against $\log(\gamma)$, for $n\leq20$.
}
\label{fig:pplots}
\end{figure}

We will now restrict to the case $m=2$ and vary the parameters $r=MR$. The integrand of
\eqref{eq:feip} has poles at
\begin{equation}
x-\frac{\alpha\tau_2^3}{8\pi^3}x^3=2n\pi i
\;,
\label{eq:5.2.7}
\end{equation}
where $n\in\mathbb{Z}$ for the R sector ($\lambda=-1$) and
$n\in\mathbb{Z}+\frac{1}{2}$ for the NS sector ($\lambda=1$). The roots of equation
\eqref{eq:5.2.7} are 
\def\ftwoone{{F\!}}
\begin{align}
   x_1(n) 
&= 2n\pi i\,
   \ftwoone\left(\tfrac{1}{3},\tfrac{2}{3};\tfrac{3}{2};-n^2\gamma\right)
\label{eq:5.2.8}
\\
&=2n\pi i
  \left(
   \tfrac{3}{2}\ftwoone\left(\tfrac{1}{3},\tfrac{2}{3};\tfrac{1}{2};1{+}n^2\gamma \right)
  {-}
   \tfrac{\sqrt{3}}{2}\sqrt{1{+}n^2\gamma}
   \ftwoone\left(\tfrac{5}{6},\tfrac{7}{6};\tfrac{3}{2};1{+}n^2\gamma\right)
  \right)
\;,
\nonumber
\\
   x_2(n)
&= - 2n\pi i
    \left(
    \tfrac{1}{2}\ftwoone\left(\tfrac{1}{3},\tfrac{2}{3};\tfrac{3}{2};-n^2\gamma\right)
   -\tfrac{1}{2n}\sqrt{-\tfrac{27}{\gamma}}
    \ftwoone\left(-\tfrac{1}{6},\tfrac{1}{6};\tfrac{1}{2};-n^2\gamma\right)
    \right) 
\label{eq:5.2.10}
\\
&= -6n\pi i\,\ftwoone\left(\tfrac{1}{3},\tfrac{2}{3};\tfrac{1}{2};1+n^2\gamma\right)
\;,
\nonumber
\\
   x_3(n)
&= -2n\pi i
    \left(
   \tfrac{1}{2}\ftwoone
   \left(\tfrac{1}{3},\tfrac{2}{3};\tfrac{3}{2};-n^2\gamma\right)
   +\tfrac{1}{2n}\sqrt{-\tfrac{27}{\gamma}}\,
   \ftwoone\left(-\tfrac{1}{6},\tfrac{1}{6};\tfrac{1}{2};-n^2\gamma\right)
   \right)
\label{eq:5.2.12}
\\
&=2n\pi i\left(\tfrac{3}{2}\ftwoone\left(\tfrac{1}{3},\tfrac{2}{3};\tfrac{1}{2};1{+}n^2\gamma\right)
{+}\tfrac{\sqrt{3}}{2}\sqrt{1{+}n^2\gamma}\,\ftwoone\left(\tfrac{5}{6},\tfrac{7}{6};\tfrac{3}{2};1{+}n^2\gamma\right)\right)
\;,
\nonumber
\end{align}
where 
\be
 \ftwoone\,(a,b;c;z) \equiv{}_2F_1(a,b;c;z)\;,\;\; \hbox{ and }\;\;
 \gamma = \frac{27 \alpha \tau_2^3}{8\pi}
\;.
\ee
It is helpful to plot these roots in the complex plane for various
values of $\gamma$, as in figure \ref{fig:pplots}.

We divide \eqref{eq:5.1.10} by 2 since we are just interested in right
movers and start with $M=0$. We can continue $M$ in the complex plane
so that it encircles the poles $x_i$ and then return $M$ to zero. When
the integration contour is then returned to the positive real axis we
pick up contributions from the poles which add $\pm ix_i$ to the
integral, the $\pm$ depends on which way we encircle the poles.

The one-particle energies we found in section \ref{sec:MT1}
are given by $i x_1(n)$ for $n>0$. 
Since $x_1+x_2+x_3$, we could get the same answer by instead using $-i x_2(n)-ix_3(n)$.

The hypergeometric function is real if its argument is real and less
than or equal to 1. Hence 
all the $i x_i$ are real if $0>\alpha>-\frac{32\pi}{27(2n+1)^2}$, but
for  $\alpha<-\frac{32\pi}{27(2n+1)^2}$ then $i x_3$ is real and 
$i x_1$ and $i x_2$ are a complex conjugate pair.

Finally we note that the roots in the upper half plane are 
$x_1(n)$, $x_2(n)$ and $-x_3(n)$ for $n>0$.

\subsection{Conjectured exact result}

We consider the simplest case where the GGE only has a non-zero chemical
potential for the charge $I_3$. In this case our conjecture is that
the modular transform of the expectation value $\vev{\exp(\alpha I_3)}$
is given by the product over terms corresponding to {\em all} of the
poles $x_i$ (\ref{eq:5.2.8}--\ref{eq:5.2.12})
in the upper half plane where $\gamma$ has now become
\be
\gamma=-\frac{27\alpha\tau^3}{8\pi i}\;.
\ee 
(Taking the pole in the upper half plane means that the energies $i x_1$ have
negative real part and the resulting expressions are convergent).

This conjecture replaces the asymptotic formulae \eqref{eq:NS+a}, \eqref{eq:R+a}
and \eqref{eq:NS-a} by
\begin{align}
\Tr_{\NS,+}\left(\hq^{L_0 - c/24} e^{\alpha I_3}\right)
&=\hat\chi^{\NS,+}(\tau,\alpha)\nonumber
\\
&= 
q^{h_0^{\NS}(\tau,\alpha)}   
\prod_{k=1/2} 
\left (1 + e^{\tau x_1(k)}  \right)
\left (1 + e^{\tau x_2(k)}  \right)
\left (1 + e^{-\tau x_3(k)}  \right)\;,
\label{eq:NS+a2}
\\
\Tr_{\R,+}\left(\hq^{L_0 - c/24} e^{\alpha I_3}\right)
&= 2^{1/2}\cdot \hat\chi^{\NS,-}(\tau,\alpha) 
\nonumber
\\
&= 2^{1/2}\cdot 
q^{h_0^{\NS}(\tau,\alpha)}
  \prod_{k=1/2} 
\left (1 - e^{\tau x_1(k)}  \right)
\left (1 - e^{\tau x_2(k)}  \right)
\left (1 - e^{-\tau x_3(k)}  \right)\;,
\label{eq:R+a2}
\\
\Tr_{\NS,-}\left(\hq^{L_0 - c/24} e^{\alpha I_3}\right)
&=
2^{-1/2} \cdot \hat\chi^{\R,+}(\tau,\alpha)  
\nonumber
\\
&= 2^{-1/2}\cdot 2\left (1 + e^{\tau x_2(0)}  \right)\,
q^{h_0^{\R}(\tau,\alpha)}
  \prod_{k=1}
\left (1 + e^{\tau x_1(k)}  \right)
\left (1 + e^{\tau x_2(k)}  \right)
\left (1 + e^{-\tau x_3(k)}  \right)\;,
\label{eq:NS-a2}
\end{align}
where the 
ground state energies $h_0^{\NS/\R}$ are defined in \eqref{eq:h0def} and
the roots $x_i(k)$ in (\ref{eq:5.2.8}--\ref{eq:5.2.12}).
Note that $-x_3(k) = x_1(k)+x_2(k) = x_2(-k)$, so there are three
different ways to write the expressions (\ref{eq:NS+a2}--\ref{eq:NS-a2}).
Also, since $x_1(0)=0$ we have $(1+x^{\tau x_1(0)})=2$ and also $x_2(0)=-x_3(0)$, which is why we only include this term once in \eqref{eq:NS-a2}.

As can easily be seen, these have the form of a trace over a space
with three sets of fermionic modes, rather than the expected one set. 

The asymptotic formulae previously found are recovered by dropping the factors
involving the roots $x_2$ and $x_3$. We note again that the
expressions
\eqref{eq:NS+a2} etc have identical asymptotic expansions to the
previous expressions \eqref{eq:NS+a} etc.

This gives predictions for the ``expectation values'' (with the
fermionic character divided off)
\begin{align}
\vev{e^{\alpha I_3}}^{\NS,+}(-1/\tau)
&= \frac{\hat\chi^{\NS,+}(\tau,\alpha)}{\chi^{\NS,+}(\tau)}
\nonumber\\
&= 
q^{h_0^{\NS}(\tau,\alpha)+1/48}   
\prod_{k=1/2} 
\frac{
  \left (1 + e^{\tau x_1(k)}  \right)
  \left (1 + e^{\tau x_2(k)}  \right)
  \left (1 + e^{-\tau x_3(k)}  \right)}{1 + q^k}
\;,\label{eq:NS+a4}
\\
\vev{e^{\alpha I_3}}^{\R,+}(-1/\tau)
&= \frac{\hat\chi^{\NS,-}(\tau,\alpha)}{\chi^{\NS,-}(\tau)}
\nonumber\\
&=
q^{h_0^{\NS}(\tau,\alpha)+1/48}
  \prod_{k=1/2} \frac{
\left (1 - e^{\tau x_1(k)}  \right)
\left (1 - e^{\tau x_2(k)}  \right)
\left (1 - e^{-\tau x_3(k)}  \right)
}{1-q^k}
\;,\label{eq:R+a4}
\\
\vev{e^{\alpha I_3}}^{\NS,-}(-1/\tau)
&=
\frac{\hat\chi^{\R,+}(\tau,\alpha)}{\chi^{\R,+}(\tau)}
\nonumber
\\
&= 
q^{h_0^{\R}(\tau,\alpha)-1/24}\left (1 + e^{\tau x_2(0)}  \right)
  \prod_{k=1}\frac{
\left (1 + e^{\tau x_1(k)}  \right)
\left (1 + e^{\tau x_2(k)}  \right)
\left (1 + e^{-\tau x_3(k)}  \right)
}{1+q^k}
\;.\label{eq:NS-a4}
\end{align}

We will now consider the leading order behavoiur of the $x_i(n)$ for large negative $\gamma$ and positive $n$. For negative $n$ we can use the relations $x_1(-n)=-x_1(n)$ and $x_2(-n)=-x_3(n)$. For large positive $z$ the hypergeometric functions have the following leading order terms
 \begin{eqnarray}
{}_2F_1\left(\frac{1}{3},\frac{2}{3};\frac{3}{2};z\right)&=&\frac{3}{4^\frac{1}{3}}e^{-\frac{\pi i}{3}}z^{-\frac{1}{3}}+O(z^{-\frac{2}{3}})\;,\\
\frac{1}{\sqrt{z}}{}_2F_1\left(-\frac{1}{6},\frac{1}{6};\frac{1}{2};z\right)&=&\frac{1}{4^\frac{1}{3}}e^{\frac{\pi i}{6}}z^{-\frac{1}{3}}+O(z^{-\frac{2}{3}})\;.
\end{eqnarray}
Using these we can find the leading order terms in the $x_i$
\begin{eqnarray}
x_1(n)&=&3(2n)^\frac{1}{3}\pi e^{\frac{\pi i}{6}}(-\gamma)^{-\frac{1}{3}}+O(\gamma^{-\frac{2}{3}})\;,\label{eq:x1large}\\
x_2(n)&=&3(2n)^\frac{1}{3}\pi e^{\frac{5\pi i}{6}}(-\gamma)^{-\frac{1}{3}}+O(\gamma^{-\frac{2}{3}})\;,\label{eq:x2large}\\
x_3(n)&=&3(2n)^\frac{1}{3}\pi e^{-\frac{\pi i}{2}}(-\gamma)^{-\frac{1}{3}}+O(\gamma^{-\frac{2}{3}})\;.\label{eq:x3large}
\end{eqnarray}
Note that these leading order terms match the pattern of the roots in
figure \ref{fig:pplots} where as $|\gamma|$ increases the roots
$x_1(n)$ converge towards the rays $e^{\frac{\pi i}{6}}$ and
$e^{-\frac{5\pi i}{6}}$, $x_2(n)$ towards $e^{\frac{5\pi i}{6}}$ and
$e^{-\frac{\pi i}{6}}$ and $x_3(n)$ towards $e^{\frac{\pi i}{2}}$ and
$e^{-\frac{\pi i}{2}}$. 

We can use (\ref{eq:x1large}--\ref{eq:x2large}) to show that for large negative $\alpha$ (equivalently large $\gamma$) the exact expressions and the conjectures match to leading order. If we take the logarithm of (\ref{eq:NS+a2}--\ref{eq:NS-a2}) then for large negative $\alpha$ we have
\begin{eqnarray}
\log\left(\text{Tr}_{NS,\pm}\left(\hat q^{L_0-c/24}e^{\alpha I_3}\right)\right)=\frac{7\alpha}{1920}+\dots\;,\label{eq:NSexactlarge}\\
\log\left(\text{Tr}_{R,+}\left(\hat q^{L_0-c/24}e^{\alpha I_3}\right)\right)=-\frac{\alpha}{240}+\dots\;,\label{eq:Rexactlarge}
\end{eqnarray}
where the $\dots$ represent terms that are suppressed for large negative $\alpha$. For the right hand side we can see that the ground state energies $h_0^{\text{NS/R}}$ both vanish as $\alpha$ becomes large and negative. Just looking at the logarithm of the product terms we can rewrite them as a sum
\begin{eqnarray}
&\log\left(\prod_{k>0}\left(1\pm e^{\tau x_1(k)}\right)\left(1\pm e^{\tau x_2(k)}\right)\left(1\pm e^{-\tau x_3(k)}\right)\right)=\nonumber\\
&\sum_{k>0}\left(\log\left(1\pm e^{\tau x_1(k)}\right)+\log\left(1\pm e^{\tau x_2(k)}\right)+\log\left(1\pm e^{-\tau x_3(k)}\right)\right)\;,
\end{eqnarray}
and if we look at each of the $x_i$ contributions separately and replace $x_i$ with it's leading $\alpha$ term then we have sums of the form
\be
\sum_{k>0}\log\left(1\pm \exp\left((2\pi)^\frac{4}{3}(-i)^\frac{1}{3}e^{i\theta_i}\left(\frac{k}{-\alpha}\right)^\frac{1}{3}\right)\right)\;,\label{eq:alphasum}
\ee
where $\theta_1=\frac{\pi}{6}$, $\theta_2=\frac{5\pi}{6}$ and $\theta_3=\frac{\pi}{2}$. If we define the variable $x_k=\frac{k}{-\alpha}$ then all $x_k$ are positive and we have the spacing $\Delta x=x_{k+1}-x_k=\frac{-1}{\alpha}$. We will rewrite \eqref{eq:alphasum} in terms of the $x_k$ to get
\be
-\alpha\sum_{k>0}\Delta x\log\left(1\pm \exp\left((2\pi)^\frac{4}{3}(-i)^\frac{1}{3}e^{i\theta_i}x_k^\frac{1}{3}\right)\right)\;.\label{eq:xsum}
\ee
In the large $\alpha$ limit the sum \eqref{eq:xsum} becomes the integral
\be
-\alpha\int_0^\infty dx\log\left(1\pm \exp\left((2\pi)^\frac{4}{3}(-i)^\frac{1}{3}e^{i\theta_i}x^\frac{1}{3}\right)\right)\;.
\ee
We use the substitution $u=-(2\pi)^{\frac{4}{3}}(-i)^\frac{1}{3}e^{i\theta_i}x^\frac{1}{3}$ where we choose the branch of $(-i)^\frac{1}{3}$ so that the real part of $u$ tends to positive infinity as $x$ goes to positive infinity. The integrand only has singularities on the imaginary axis in the $u$ plane and hence we can move the integration contour onto the positive real axis without picking up any poles (the additional arc contour that connects the two lines vanishes exponentially as we take the lines to infinity). If we do integration by parts on the integral we then arrive at
\be
\frac{3i\alpha e^{-3i\theta_i}}{(2\pi)^4}\int_0^\infty du\;u^2\log(1\pm e^{-u})=\pm\frac{i\alpha e^{-3i\theta}}{(2\pi)^4}\int_0^\infty du\frac{u^3}{e^u\pm 1}=\begin{cases}-\frac{i\alpha}{240}e^{-3i\theta_i} &-\\
\frac{7i\alpha}{1920}e^{-3i\theta_i}&+\end{cases}\;.
\ee
We can sum over the $i$ to find
\be
i\sum_{i=1}^3e^{-3i\theta_i}=1\;,
\ee
and finally this gives us the leading large $\alpha$ term for the conjecture
\begin{eqnarray}
\log\left(\hat\chi^{\text{NS/R},+}\right)=\frac{7\alpha}{1920}+\dots\;,\\
\log\left(\hat\chi^{\text{NS},-}\right)=-\frac{\alpha}{240}+\dots\;,
\end{eqnarray}
where the $\dots$ are suppressed terms. These leading terms match
those in \eqref{eq:NSexactlarge} and \eqref{eq:Rexactlarge}
and this provides some non-trivial analytic support for our conjecture.

\section{Numerical evidence}
\label{sec:numerical}

Our evidence for the correctness of the formulae \eqref{eq:NS+a2},
\eqref{eq:R+a2} and \eqref{eq:NS-a2} is based mainly on the numerical
agreement (to within numerical accuracy) across a range of $\tau$ and
$\alpha$.
In table \ref{tab:1} we give the values of the modular transforms of
our exact results for the expectation values
$ \vev{ e^{\alpha I_3} }^{\NS/\R \pm}(-1/\tau)$, 
the conjectured exact expressions \eqref{eq:NS+a4}--\eqref{eq:NS-a4},
and the asymptotic expressions from \eqref{eq:NS+a}--\eqref{eq:NS-a} 
(which we denote $\hat\chi_{\NS/\R\pm}^{(1)}(\tau,\alpha)/\chi_{\NS/\R\pm}(\tau)$ since it is
given by the same form as the conjectured exact expression by only including the roots $x_1(k)$).

{
\captionsetup{margin=10pt}
\begin{table}
\centering
\begin{tabular}[htb]{|l|l|l|l|l|}
\hline
$\alpha$ &\multicolumn{4}{c|}{$\tau=i$}
\\\hline
-0.01
& 0.9999909023
& 0.9999909023 
& 0.9999909023
& 0.9999909023
\\
& 
&$\; -\, 5.1697517 \times 10^{-126} i$ 
&
&
\\\hline
  -0.1 
& 0.9990978822 
& 0.9990978822 
& 0.9990978822 
& 0.9990978822
\\
&
&$\; + \,1.0567308 \times 10^{-13} i $
& 
& 
\\\hline
-1
& 0.9914336696 
& 0.9914091375 
& 0.9914336280
& 0.9914336696
\\
&
&$\; - \, 0.000044605\, i$
&
&
\\ \hline
-10
& 0.9356250621 
& 0.9374288339 
& 0.9326249941
& 0.9356250621 
\\
&
&$\; +\,  0.0240515109\, i$
& 
& 
\\
\hline
& $\vev{e^{\alpha I_3}}^{NS+}(-1/\tau) $
& ${\hat\chi^{(1)}_{\NS+}}/{\chi_{\NS+}}(\tau) $
& ${\hat\chi^{(1,2)}_{\NS+}}/{\chi_{\NS+}}(\tau) $
& ${\hat\chi_{\NS+}}/{\chi_{\NS+}}(\tau) $
\\ \hline
\multicolumn{5}{c}{}
\\
\hline
$\alpha$ &\multicolumn{4}{c|}{$\tau=(-1+4i)/3$}
\\\hline
-0.1 
& 1.0008014459 
& 1.0008014459 
& 1.0008014459 
& 1.0008014459 \\
&$\; +\, 0.0012333495\,i $
&$\; +\, 0.0012333495\,i $
&$\; +\, 0.0012333495\,i $
&$\; +\, 0.0012333495\,i $\\ \hline
-1
& 1.0076484482
& 1.0076114564
& 1.0076484114
& 1.0076484482 \\
&$\; +\, 0.009370977\, i $
&$\; +\, 0.009397878\, i $
&$\; +\, 0.009370902\, i $
&$\; +\, 0.009370977\, i $\\\hline
-10
& 1.03116010611
& 1.01844820142
& 1.02869050956
& 1.03116010611 \\
&$\; +\, 0.0490164216\, i $
&$\; +\, 0.0639418298\, i $
&$\; +\, 0.0482967963\, i $
&$\; +\, 0.0490164216\, i $
\\\hline
& $\vev{e^{\alpha I_3}}^{NS-}(-1/\tau) $
& ${\hat\chi^{(1)}_{\R+}}/{\chi_{\R+}}(\tau) $
& ${\hat\chi^{(1,2)}_{\R+}}/{\chi_{\R+}}(\tau) $
& ${\hat\chi_{\R+}}/{\chi_{\R+}}(\tau) $
\\ \hline
\end{tabular}
\caption{Table of values of a GGE including a coupling to a single
  charge $I_3$ in two different sectors calculated using the exact
  result from the expression of the charge as a fermion bilinear in
  one channel and the
  asymptotic expression, an improved asymptotic expression and a
  conjectured exact answer in the opposite channel}
\label{tab:1}
\end{table}
}

We also include a modified guess 
(which we denote $\hat\chi_{\NS/\R\pm}^{(1,2)}(\tau,\alpha)/\chi_{\NS/\R\pm}(\tau)$)
including the product over only the
roots $x_1(k)$ and $x_2(k)$ which is the minimal change required to
make the expressions real and single-valued for real $\alpha$ and purely
imaginary $\tau$.

For large negative values of $\alpha$, the roots appear to become
dense, requiring a very large number of terms to be included in the
products (greater than 1000) to achieve reliable numerical results.

\section{Conclusions}
\label{sec:conc}

We have considered the torus partition function for the free fermion
model coupled to KdV charges, or equivalently the partition function
of generalised Gibbs ensembles. 
Using the asymptotic expansion of the generalised Gibbs ensemble about
$\alpha=0$ allowed us to find an asymptotic expression for the modular
transform of the GGE. However the first step involved expanding the
GGE as an asymptotic series with vanishing radius of convergence so it
is not clear how the above method could be adapted to find the exact
from of the modular transformed GGE. 
The asymptotic expressions take the form of a trace over a fermion Fock
space, but if any of the chemical potentials of the higher charges are
non-zero then the expression must be wrong. This is most easily seen
by looking at a rectangular torus and noting that the original GGE is
real but the asymptotic form in the crossed channel is not.

We also analysed the problem from the TBA point of view 
(in the
particular case of coupling to the single charge $I_3$ and for the
modular transform $S:\tau \to -1/\tau$),
to see if
the spectrum predicted by TBA was the same as that found from our
asymptotic analysis. 
Our TBA analysis confirmed the energies of the modes we found in the
asymptotic analysis but revealed two extra sets of poles which could contribute
to the spectrum. One set of these poles solves the reality problem of
the asymptotic expression but still gives an expression which is not correct. Including both
extra sets, however, gives an expression which agrees numerically with
the exact result (in the opposite channel) to a high degree of
precision and which we conjecture is the exact result.

The conjectured result for the modular transform of a GGE takes the
form of a trace over 
three 
copies of the free-fermion Fock space. 
The interpretation of this result is not clear to us. This could be an
artifact, the ``extra'' copies really being a change to the ground
state energy; it could instead indicate that there is a fundamental change to
the nature of the Hilbert space in the crossed channel caused by the
introduction of a coupling to the charge $I_3$.

We have also, so far, only found a conjectured closed form expression for the
action of $S$; it remains to prove the conjecture
(it could be wrong despite passing numerical and analytic tests)  
and to extend it to arbitrary elements of the modular group.

This is reminiscent of the spectrum of other perturbed systems which
exhibit a similar instability such as the spectrum of the boundary
Yee-Lang model perturbed by a large boundary magnetic field
\cite{DPTW} (where one copy of a Virasoro representation splits into
three copies, one real set and two with complex conjugate energies)
and the spectrum of the non-Hermitian spin chain
\cite{16}. 
This has sometimes been interpreted as the result of a
phase-transition caused by the instability of the vacuum to the
perturbation, something which might also be occurring here.
This clearly deserves more study and a better understanding.

In this paper we have only investigated in detail the GGE with one coupling
to a single charge. This needs to be extended to several charges and to an
infinite set of charges.

It will also be worth investigating the large $\alpha$ limit in which
the roots appear to condense onto cuts in the complex plane.

Finally, it is worth looking at other models 
(such as the Lee-Yang model)
for which, at present,
there is no closed form for the GGE in either channel to see if any
evidence for similar behaviour can be found.

\subsubsection*{ Acknowledgments}

We would like to thank B.~Doyon, A.~Dymarsky, S.~Murthy, S.~Ross, and G.~Tak\'acs for helpful
discussions at various times. We would also like to thank the JHEP referee for careful reading of the manuscript.

MD would like to thank EPSRC for support under  grant EP/V520019/1.

\appendix

\section{Appendix: The free fermion and the Ising model}
\label{app:ff}

%

Firstly, we define the operator $(-1)^F$ which squares to 1 and which 
anticommutes with all fermion modes,
\be
 \{\psi_m,(-1)^F\} = \{\bar\psi_m,(-1)^F\}=0
\;.
\ee
States with eigenvalue $+1$ are called bosonic and those with
eigenstate $-1$ are fermionic.

The spaces $\cH_\NS$ and $\cH_\R$ are defined as follows.

\subsection{The NS sector}

We define the NS vacuum $\vac_\NS$ by
\be
  \psi_m \vac_\NS = 0\;,\;\; m>0
\;,\;\;\;\;
  (-1)^F\vac_\NS = \vac_\NS\;,
\ee
and the space $\cH_\NS$ has basis
\be
  \{ \psi_{-m_1}\ldots\psi_{-m_n}\vac_\NS\;,\;\; m_1 > m_2  \ldots >
  m_n \geq \tfrac 12 \}
\;.
\ee
In the NS sector, $L_0$ takes the form
\be
 L_0 = \sum_{m>0} m \,\psi_{-m}\psi_m
\;.
\label{eq:L0NS}
\ee
The space $\cHNS$ splits into two irreducible representations of the
Virasoro algebra with $h=0$ and $h=1/2$ given by the states which have
an even or an odd number of fermion modes, respectively.
If we define the traces
\begin{align}
  \Tr_{\NS,+}(X) = \Tr_{\cH_\NS}(X)
\;,\;\;\;\;\;\;
  \Tr_{\NS,-}(X) = \Tr_{\cH_\NS}(\,(-1)^F\,X)
\;,\;\;
\label{eq:NSTrdef}
\end{align}
then we can define
\begin{align}
  \chi^{\NS,+} 
&= \Tr_{\NS,+} ( q^{L_0 - c/24} )
= q^{-1/48} \prod_{n=0}^\infty (1 + q^{n+1/2})
\;,\;\;\label{eq:NS+def}
\\
  \chi^{\NS,-} 
&= \Tr_{\NS,-}( q^{L_0 - c/24})
= q^{-1/48} \prod_{n=0}^\infty (1 - q^{n+1/2})
\;.
\label{eq:NS-def}
\end{align}
These are related to the characters of the irreducible Virasoro
representations by
\begin{align}
  \chi_0 &= \Tr_{\NS}(\, \tfrac{(1 + (-1)^F)}2 q^{L_0 - c/24})
\;,\;\;  
  &\chi^{\NS,+} &= \chi_0 + \chi_{1/2}
\;,\;\;
\\
  \chi_{1/2} &= \Tr_{\NS}(\, \tfrac{(1 - (-1)^F)}2 q^{L_0 - c/24})
\;,\;\;
  &\chi^{\NS,-} &= \chi_0 - \chi_{1/2}
\;.
\end{align}

\subsection{The R sector}

We define the R vacuum $\vac_R$ by
\be
  \psi_m \vac_\R = 0\;,\;\; m>0
\;,\;\;\;\;
  (-1)^F\vac_\R = \vac_\R\;,
\ee
and the space $\cHR$ has basis
\be
  \{ \psi_{-m_1}\ldots\psi_{-m_n}\vac_\R\;,\;\; m_1 > m_2  \ldots >
  m_n \geq 0 \}
\;.
\ee
In the R sector, $L_0$ takes the form
\be
 L_0 = \sum_{m>0} m \,\psi_{-m}\psi_m + \tfrac 1{16}
\;.
\label{eq:L0NS}
\ee
Note that the space of states with $h=1/16$ is two-dimensional and spanned
by the bosonic state $\vac_\R$ and the fermionic state $\psi_0\vac_R$.

Correspondingly, 
The space $\cHR$ splits into two copies of irreducible representations of the
Virasoro algebra with $h=1/16$ formed of the bosonic and the fermionic
states respectively.
If we define the traces
\begin{align}
  \Tr_{\R,+}(X) = \Tr_{\cHR}(X)
\;,\;\;\;\;\;\;
  \Tr_{\R,-}(X) = \Tr_{\cHR}(\,(-1)^F\,X)
\;,\;\;
\label{eq:RTrdef}
\end{align}
then we can define
\begin{align}
  \chi^{\R,+} 
&= \Tr_{\R,+} ( q^{L_0 - c/24} )
= 2\,q^{1/24} \prod_{n=1}^\infty (1 + q^{n})
= q^{1/24} \prod_{n=0}^\infty (1 + q^{n})
\;,\;\;\label{eq:R+def}
\\
  \chi^{\R,-} 
&= \Tr_{\R,-}( q^{L_0 - c/24})
= 0
\;.
\label{eq:R-def}
\end{align}
These are related to the characters of the irreducible Virasoro
representations by
\be
  \chi^{\R,+} = 2\, \chi_{1/16}
\;.
\ee

\subsection{Modular properties}
The modular properties of the Virasoro characters are well known. The action
of $T$ is obvious, $\chi(\tau+1) = e^{2\pi i(h-c/24)} \chi(\tau)$.
The action of $S$ is
\be
\bpm 
 \chi_0(\tq) \\ \chi_{1/2}(\tq) \\ \chi_{1/16}(\tq)
\epm
= \bpm 
  \tfrac 12 & \phantom{-}\tfrac 12 & \phantom{-}\tfrac 1{\sqrt 2} \\
  \tfrac 12 & \phantom{-}\tfrac 12 & -\tfrac 1{\sqrt 2} \\
  \tfrac 1{\sqrt 2} & -\tfrac 1{\sqrt 2} & \phantom{-}0
\epm
\bpm 
 \chi_0(q) \\ \chi_{1/2}(q) \\ \chi_{1/16}(q)
\epm
\ee
The modular invariant combination
\be
  Z = |\chi_0|^2 + |\chi_{1/2}|^2 + |\chi_{1/16}|^2
\;,
\ee
is the partition function of the (purely bosonic) Ising model.

The analysis of the modular properties of the free-fermion model is
dictated by the periodicities of the fermion around the two cycles
of the torus. We label these by the periodicity on the torus as $A$
and $P$ for anti-periodic and periodic.
If we calculate the torus correlation functions as a trace over the
states on a cylinder, then the $A$ sector is given by a trace over the
NS sector and the $P$ sector by a trace over the R sector. 
We also need to account for the periodicity along the cylinder - and
we do this by the insertion of the operator $(-1)^F$ for the $P$
sector but no insertion for the $A$ sector.

If we denote the partition function with periodicities $R$ and $S$ by $Z_{R,S}$, then
\be
  Z_{A,A} = | \chi^{\NS,+}|^2
\;,\;\;
  Z_{A,P} = | \chi^{\NS,-}|^2
\;,\;\;
  Z_{P,A} = \tfrac 12 | \chi^{\R,+}|^2 = 2 |\chi_{1/16}|^2
\;,\;\;
  Z_{P,P} =0
\;.\;\;
\ee
The modular transforms under $T$ are
\be
  \chi^{\NS,+}(\tq) = \chi^{\NS,+}(q)
\;,\;\;
  \chi^{\R,+}(\tq) = \sqrt 2 \chi^{\NS,-}(q)
\;,
\ee 
and the partition function of the Ising model is the modular invariant
combination
\be
  Z = \tfrac 12 ( Z_{A,A} + Z_{A,P} + Z_{P,A} + Z_{P,P})
\;.
\ee

\section{Appendix: Modular forms}
\label{app:mf}

In this appendix we will list the relevant facts about modular forms that appear in this paper. Proofs of the following statements can be found in \cite{Zagier} and most of the notation will be the same. 

The full modular group will be denoted by
\be
\Gamma_1 = \text{SL}(2,\mathbb{Z})\;,
\ee
and the principal congruence subgroup $\Gamma(N)$ is defined to be the subgroup of $\Gamma_1$ such that
\be
\Gamma(N)=\{\gamma\in\Gamma_1|\gamma\equiv \text{Id}_2 (\text{mod }N)\}\;.
\ee
Consider a given matrix 
$\big(\begin{smallmatrix}  a&b\\c&d\end{smallmatrix}\big)\in\Gamma$, 
where $\Gamma$ for us is either $\Gamma_1$ or $\Gamma(2)$. If a
holomorphic function $f(\tau)$, defined in the upper half plane, has
the following transformation property 
\be
f\left(\frac{a\tau+b}{c\tau+d}\right)=(c\tau+d)^kf(\tau)\;,
\ee
then we say that the function is a holomorphic modular form of weight
$k$ on $\Gamma$. We will denote the space of modular forms of weight
$k$ on $\Gamma$ by $M_k(\Gamma)$. 

The group $\Gamma_1$ is finitely generated by the matrices
\be
\bpm1&1\\0&1\epm\;,\;\;\;\bpm0&1\\-1&0\epm\;,
\ee 
hence we only need to check that a function transforms as a modular form under
\be
T:\tau\mapsto\tau+1\;\;\;S:\tau\mapsto\frac{-1}{\tau}\;,
\ee
to verify it is an element of $M_k(\Gamma_1)$. Similarly the subgroup $\Gamma(2)$ is finitely generated by the matrices
\be
\bpm1&2\\0&1\epm\;,\;\;\;\bpm1&0\\2&1\epm\;,\;\;\;-\text{Id}_2\;,
\ee
hence we only need to check that a function transforms as a modular form under
\be
\tilde T:\tau\mapsto\tau+2\;\;\;\tilde S:\tau\mapsto\frac{\tau}{2\tau+1}\;,
\ee
to verify it is an element of $M_k(\Gamma(2))$. An important fact about the spaces $M_k(\Gamma)$ is that they are finite dimensional. The space $M_{2k}(\Gamma_1)$ is generated by the Eisenstein series (see below) and the space $M_{2k}(\Gamma(2))$ is generated by the Jacobi theta functions (see \cite{Zagier} for definitions of the Jacobi theta functions).

The Eisenstein series $E_{2k}(\tau)$ are elements of $M_{2k}(\Gamma_1)$ for $k=2,3,\dots$ and they are defined by
\be
E_{2k}(\tau)=1+\frac{2}{\zeta(1-2k)}\sum_{n=0}^\infty\frac{n^{2k-1}q^n}{1-q^n}\;,\;\;\;q=e^{2\pi i\tau}\;.
\ee
For $k=1$ the Eisenstein series $E_2(\tau)$ is quasi modular which means that under a modular transform we have the transformation property
\be
E_{2}\left(\frac{a\tau+b}{c\tau+d}\right)=(c\tau+d)^2E_2(\tau)-\frac{6i}{\pi}c(c\tau+d)\;.
\ee
Additionally we want to consider the Eisenstein series with arguments $2\tau$ and $\frac{1}{2}\tau$. By using the fact that $\tilde T=T^2$ and $\tilde S=(TST)^2$ we can see that for $k=2,3,\dots$ we have $E_{2k}(2\tau), E_{2k}(\tau), E_{2k}(\frac{1}{2}\tau)\in M_{2k}(\Gamma(2))$.

We also encounter quasi-modular forms. For our purpose we will define the space of quasi-modular forms of weight $k$ and depth $p$, denoted by $\tilde  M_k^{(\leq p)}(\Gamma)$, to be
\be
\tilde M_k^{(\leq p)}(\Gamma)=\bigoplus_{r=0}^pM_{k-2r}(\Gamma)\cdot E_2^r
\ee 

Finally we define the Serre derivative. The Serre derivative acting on a modular form $f(\tau)$ of weight $k$ is defined to be
\be
D_rf(\tau)=\frac{1}{2\pi i}\frac{d}{d\tau}f(\tau)-\frac{r}{12}E_2(\tau)f(\tau)\;.
\ee
By using the transformation of $\frac{d}{d\tau}$ under a modular transform we can see that $D_kf(\tau)$ is a modular form of weight $k+2$.


\end{document}